\documentclass[fleqn,usenatbib]{mnras}
\usepackage{newtxtext,newtxmath}
\usepackage[T1]{fontenc}
\usepackage{graphicx}	
\usepackage{amsmath}	
\usepackage{multirow}
\usepackage{pdflscape}
\usepackage{arydshln}
\usepackage{threeparttable}
\usepackage{epstopdf}
\usepackage{verbatim}
\usepackage{ulem}
\usepackage{subfigure}
\usepackage{float}
\usepackage{stfloats}
\usepackage{booktabs}
\usepackage{lineno}

\newcommand{\tSB}{t_{\rm SB}}
\newcommand{\sbd}{30\,Dor\,C}

\newcommand{\cxo}{\sl Chandra\rm}
\newcommand{\xmm}{\sl XMM-Newton\rm}
\newcommand{\rosat}{\sl ROSAT\rm}

\title[Thermal X-rays of \sbd\ West]{Thermal X-ray Emission in the Western Half of the LMC Superbubble \sbd}

\author[Y. -H. Chi et al.]{
Yi-Heng Chi,$^{1}$
Han-Xiao Chen,$^{1}$
Yang Chen,$^{1,2}$\thanks{E-mail: ygchen@nju.edu.cn}
Yi-Fan Meng,$^{1}$
Ping Zhou,$^{1,2}$\thanks{E-mail: pingzhou@nju.edu.cn}
Lei Sun,$^{1,2}$
and Wei Sun$^{3,4}$ 
\\
$^{1}$School of Astronomy and Space Science, Nanjing University, Nanjing, 210023, People’s Republic of China\\
$^{2}$Key Laboratory of Modern Astronomy and Astrophysics, Ministry of Education, Nanjing, 210023, People’s Republic of China\\
$^{3}$Purple Mountain Observatory, Chinese Academy of Sciences, Nanjing 210008, China \\
$^{4}$Key Laboratory of Dark Matter and Space Astronomy, Chinese Academy of Sciences, Nanjing 210008, China
}
 
\date{Accepted XXX. Received YYY; in original form ZZZ}

\pubyear{2024}

\begin{document}
\label{firstpage}
\pagerange{\pageref{firstpage}--\pageref{lastpage}}
\maketitle
\begin{abstract}

While \sbd\ is a unique superbubble in the Large Magellanic Cloud for its luminous non-thermal X-ray emission, the thermal X-ray emission it emanates has not yet been thoroughly investigated and well constrained. Based on the separate $\sim1$\,Ms deep \xmm\ and \cxo\ observations, we report the discovery of the thermally-emitting plasma in some portions of the western half of \sbd. The thermal emission can be reproduced by a collisional-ionization-equilibrium plasma model with an average electron temperature of $\sim0.4$\,keV. We find a significant overabundance of the intermediate-mass elements such as O, Ne, Mg, and Si, which may be indicative of a recent supernova explosion in \sbd. Dynamical properties in combination with the information of the OB association LH~90 suggest that the internal post-main-sequence stars dominate the power of the superbubble and blow it out in the past $\sim1$\,Myr.

\end{abstract}




\section{Introduction}
Superbubbles (SBs) are large shell-like structures of $10^2$--$10^3$\,pc in diameter, composed of low-density, hot ($\sim10^6\,\rm{K}$) shock-heated gas \citep{MacLaw1988}. In SBs, both stellar winds from massive stars in OB associations and supernovae (SNe) may jointly play crucial physical roles.
Energetic stellar winds of massive stars and intense shock waves of SNe sweep up the surrounding interstellar medium (ISM) and ejecta, resulting in a thin, dense, cold shell. As a result of the low density ($\sim0.01\,\rm cm^{-3}$) inside these bubbles, the shock wave needs to travel a long distance before it significantly decelerates, so effective particle acceleration occurs on a longer timescale than for single/isolated supernova remnants \citep[SNRs,][]{Lopez2020}. At the same time, due to the contribution of both massive stars' winds and SNe to the total SB energy, the SB stores more energy than a single SNR and thus becomes a possible origin of high-energy cosmic rays. However, detailed studies of Galactic SBs are complex as they require observations with a large field of view. The interference of foreground or background objects along the line of sight and strong interstellar extinction at the Galactic plane also affect observations in several wavebands, including optical, ultraviolet, and X-ray.

The Large Magellanic Cloud (LMC) is an ideal laboratory for the study of various astrophysical objects and their physical mechanisms, with favorable conditions such as close proximity \citep[at a distance $\sim50$\,kpc,][]{Feast1999}, low inclination \citep[$\sim$ 35°, almost face-on,][]{VanMarel2001}, and low foreground absorption \citep[$A_v$ < 0.3\,mag,][]{Bessel1991}. In addition, the LMC harbors a large number of SBs, SNRs, pulsar wind nebulae (PWNe), and various other high-energy astrophysical objects that can be resolved by modern observational instruments.

30 Doradus C (\sbd\ for short) is one of the noticeable SBs in the LMC, located to the south-west of the star formation region 30 Doradus (30 Dor). It was discovered by \citet{LeMarne1968} through radio continuum observation. \citet{Mills1984} figured out a shell-like structure $\sim3'$ or $\sim40$\,pc in radius in the 843 MHz band. After that, it was detected in the X-ray band by {\sl Einstein} for the first time \citep{Long1981} and was detected as a shell structure by \rosat\ \citep{Dunne2001}. In 2000, \xmm's first light observation was aimed at it, through which \citet{Dennerl2001} found a complete ring-like structure with $\sim6'$ in diameter in soft and hard X-rays.

X-ray emissions are widely observed from shocked gas and central OB associations in SBs, such as RCW~38 \citep{Wolk2002} and Westerlund~1 \citep{Muno2006} inside our Galaxy, N\,11 \citep{Maddox2009}, N\,51D \citep{Cooper2004}, and \sbd\ \citep{Bamba2004, Smith2004, Yamaguchi2009, Kavanagh2015} in the LMC, and IC\,131 \citep{Tullmann2009} in M\,33. Among the SBs in the Local Group, only \sbd\ has a non-thermal X-ray shell along with the brightest non-thermal X-ray and TeV emission \citep{H.E.S.S2015}, which has made it an ideal laboratory to study the mechanism of non-thermal emission of SBs. Non-thermal emission is distributed throughout the SB and it is brighter in the western shell than in the eastern shell. However, the sign of thermal X-rays emerges among the predominant sea of non-thermal photons. By \cxo\ and \xmm\ observations, \citet{Bamba2002, Bamba2004} revealed the differences in the spatial distribution between thermal and non-thermal emission in \sbd, discovering that while the non-thermal emission is enhanced in the north and west, the thermal emission might mainly concentrate on the south-eastern part. 

Previous researches on \sbd\ have mostly been focused on the non-thermal properties, but the thermal plasma still lacks detailed studies. Based on the X-ray analysis of the shell and the full coverage of the interior, it is generally conjectured that there could be thermal components in the eastern region, while the spectra in the western region can be well fitted by a single non-thermal model \citep{Kavanagh2015, Babazaki2018}. A newly identified SNR MCSNR J0536-6913 projectively at the south-east of the SB's shell is associated with the most luminous thermal emission but is suggested to be outside of \sbd\ \citep{Kavanagh2015}. The existence of diffuse thermal emissions in the eastern half of \sbd\ is generally confirmed by previous studies despite differences in properties like temperatures \citep{Bamba2002, Kavanagh2015, Babazaki2018, Lopez2020, Sasaki2022}. On the other hand, in the western part of \sbd\ (hereafter \sbd\ West), only \citet{Lopez2020} and \citet{Sasaki2022} declared the detection of thermal emissions; but both studies lacked a local background, so that the reported emissions may suffer substantial line-of-sight contamination.

However, thermal emission could not only provide basic physical parameters such as plasma temperature, density, and metal abundance of \sbd\ but also benefit us in revealing its evolution related to internal bluish stars and exploded SNe. Therefore, the thermal emission in the \sbd\ is worth an in-depth study. Our work will mainly focus on the \sbd\ West in view of the lack of solid evidence of thermal component here in previous works, and the primary scientific objective of our work is to search for it using \cxo\ and \xmm\ observations and to reveal signatures of feedback from SNe and massive stars.

This paper is organized as follows: in Section \ref{sec:data}, we present our data reduction of the \cxo\ and \xmm\ observations in detail. The results of our study are divided into three parts in Section \ref{sec:res}: the spectra and likely associations of point-like sources in Section \ref{sec:pls}, our spectral fittings in Section \ref{sec:spec} and narrow-band maps to outline emission lines in Section \ref{sec:flux}. Later, we discuss the thermal emission properties of the SB in Section \ref{sec:prop} and the dynamical properties in Section \ref{sec:mo}. The work is summarized in Section \ref{sec:sum}.

\section{Data reduction}
\label{sec:data}
\subsection{Chandra}

We retrieved \cxo\ data on SB \sbd\ for spatially resolved spectroscopic analysis taking advantage of \cxo's high angular resolution. As there are only 5 observations (ObsIDs 17904, 19925, 20339, 21949, 22006) with a total exposure less than 170\,ks specifically proposed for \sbd, we used the \it find\_chandra\_obsid \rm command in the \cxo\ Interactive Analysis of Observations \citep[CIAO,][version 4.14 and CALDB version 4.9.7]{ciao} software to search for other observations in which the Advanced CCD Imaging Spectrometer (ACIS) also covers the SB. We selected 31 observations (shown in Table \ref{tab:chandraobs}) with individual exposure $\gtrsim10$\,ks and a total exposure time $\gtrsim1.1$\,Ms. All the data were reprocessed with \it chandra\_repro \rm script. 

\begin{table}
    \setlength\tabcolsep{4pt}
    \centering
    \caption{\cxo\ observations}
    \label{tab:chandraobs}
    \begin{tabular}{cccccc}
    \hline
        Observation  & Observation & PI & Instrument & Exposure  & Target \\
        ID & date  &  & time (ks) &  \\ \hline
        124 & 1999-10-06 & Garmire & ACIS-S & 47.33 & SN 1987A \\ 
        1044 & 2001-04-25 & Garmire & ACIS-S & 17.18 & SN 1987A \\ 
        1387 & 1999-10-06 & Garmire & ACIS-S & 68.89 & SN 1987A \\ 
        7590 & 2007-04-17 & Canizares & ACIS-S & 35.55 & SN 1987A \\ 
        7620 & 2007-09-09 & McCray & ACIS-S & 34.62 & SN 1987A \\ 
        7621 & 2007-09-11 & McCray & ACIS-S & 36.95 & SN 1987A \\ 
        8487 & 2007-03-24 & Canizares & ACIS-S & 28.67 & SN 1987A \\ 
        8488 & 2007-03-29 & Canizares & ACIS-S & 31.66 & SN 1987A \\ 
        8543 & 2007-03-27 & Canizares & ACIS-S & 30.66 & SN 1987A \\ 
        8544 & 2007-03-28 & Canizares & ACIS-S & 19.12 & SN 1987A \\ 
        8545 & 2007-03-31 & Canizares & ACIS-S & 20.46 & SN 1987A \\ 
        8546 & 2007-04-01 & Canizares & ACIS-S & 30.64 & SN 1987A \\ 
        9580 & 2007-09-07 & McCray & ACIS-S & 34.59 & SN 1987A \\ 
        9581 & 2007-09-04 & McCray & ACIS-S & 44.96 & SN 1987A \\ 
        9582 & 2007-09-05 & McCray & ACIS-S & 44.17 & SN 1987A \\ 
        9589 & 2007-09-14 & McCray & ACIS-S & 39.53 & SN 1987A \\ 
        9590 & 2007-09-16 & McCray & ACIS-S & 24.65 & SN 1987A \\ 
        9591 & 2007-09-12 & McCray & ACIS-S & 12.85 & SN 1987A \\ 
        9592 & 2007-09-12 & McCray & ACIS-S & 12.87 & SN 1987A \\ 
        16201 & 2014-07-21 & Townsley & ACIS-I & 58.39 & 30 Dor \\ 
        16202 & 2014-08-19 & Townsley & ACIS-I & 65.13 & 30 Dor \\ 
        16203 & 2014-09-02 & Townsley & ACIS-I & 41.42 & 30 Dor \\ 
        16640 & 2014-07-24 & Townsley & ACIS-I & 61.68 & 30 Dor \\ 
        17312 & 2014-08-22 & Townsley & ACIS-I & 44.90 & 30 Dor \\ 
        17413 & 2014-09-08 & Townsley & ACIS-I & 24.65 & 30 Dor \\ 
        17414 & 2014-09-13 & Townsley & ACIS-I & 17.32 & 30 Dor \\ 
        17904 & 2017-05-03 & Kavanagh & ACIS-S & 40.47 & \sbd\ \\ 
        19925 & 2017-05-12 & Kavanagh & ACIS-S & 40.57 & \sbd\ \\
        20339 & 2018-12-23 & Lopez & ACIS-I & 25.74 & \sbd\ \\ 
        21949 & 2018-12-21 & Lopez & ACIS-I & 27.71 & \sbd\ \\ 
        22006 & 2018-12-19 & Lopez & ACIS-I & 37.58 & \sbd\ \\ \hline
        Totals & ~ & ~ & ~ & 1100.91 & ~ \\ \hline
    \end{tabular}
\end{table}

\begin{figure*}
    \centering
    \subfigure{\includegraphics[width=0.497\textwidth]{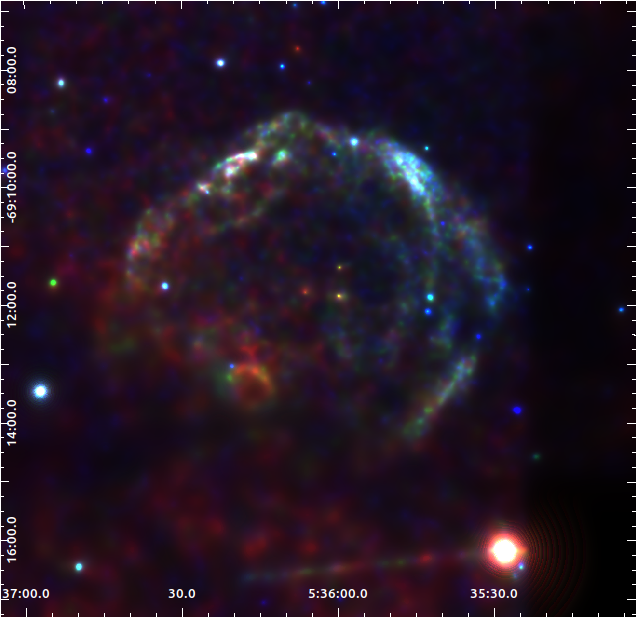}\label{fig:rgb}}
    \subfigure{\includegraphics[width=0.49\textwidth]{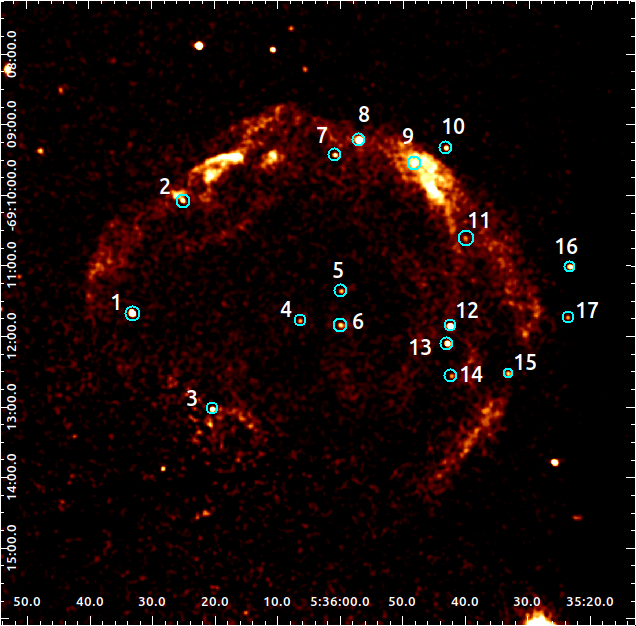}\label{fig:pls}}    
    \caption{The left panel is the adaptively smoothed, vignetting corrected \cxo\ pseudo-color image in soft (0.5--1.2\,keV), medium (1.2--2.0\,keV), and hard (2.0--7.0\,keV) bands. Non-thermal emission dominates the radiation in bluish \sbd\ West while the western half is relatively dim in the soft band.
    The bright spot to the south-west of the SB is SN\,1987A, and the shell-like MCSNR J0536-6913 is projected in the south-east of the SB. The right panel shows \sbd\ in broad (0.5--7.0\,keV) band with 17 point-like sources studied in this article circled in cyan. The image is smoothed in a scale of 3 pixels ($\sim1.5''$).}
\end{figure*}

In addition, the \it reproject\_obs \rm and \it flux\_obs \rm tasks were used to produce images, exposure maps, and point-spread function (PSF) maps for further point-like source detection. Fig. \ref{fig:rgb} shows a vignetting corrected deep RGB image of \sbd. On account of a good PSF, only 5 on-axis observations (ObsIDs 17904, 19925, 20339, 21949, 22006) were used for point-like source analysis. We employed the \it wavdetect \rm script to search the merged images in the soft (0.5--1.2\,keV), medium (1.2--2.0\,keV), hard (2.0--7.0\,keV), and broad (0.5--7.0\,keV) bands, respectively. For point-like sources, only the sources in or close (at an angular distance from the shell $< 0.5'$) to the SB were retained. Moreover, they must be detected in both the broad-band and at least one of the three sub-bands at the same time. Fig. \ref{fig:pls} shows the broad-band image with point-like sources marked in cyan.

\subsection{XMM-Newton}
\xmm\ provides us with a larger effective area, especially in low energy bands as well as a slightly higher spectral resolution than \cxo, which is especially beneficial to the spectral analysis. We obtained all the available \xmm\ observations that fully cover SB \sbd\ with each individual exposure $>20$\,ks. The total flare-filtered exposure of all 16 observations spanning over 20\,yr, as shown in Table \ref{tab:xmmobs}, amounts nearly to 1\,Ms. We processed and analyzed the data by employing the \xmm\ Science Analysis System \citep[SAS, version 19.1.0][]{sas} along with the \xmm\ Extended Source Analysis Software (XMM-ESAS, version 0.11.2) for background modeling \citep{Snowden2004}. All the data files were reprocessed with scripts {\it emchain} and {\it epchain} and then filtered with \it mos-filter \rm and \it pn-filter \rm to remove the time intervals significantly contaminated by soft proton flares. 

\begin{table}
    \setlength\tabcolsep{5pt}
    \centering
    \caption{\xmm\ observations. All observations target SN 1987A except 0115740201 (LMC) and 0113020201 (PSR J0537-6910) but cover the whole field of \sbd. The exposures are flare-filtered.}
    \label{tab:xmmobs}
    \hspace{-15pt}
        \begin{tabular}{cccccc}
        \hline
            Observation & Observation & PI & \multicolumn{3}{c}{Exposure (ks)} \\ 
             ID & date &  & pn & MOS1 & MOS2 \\ \hline
            0115740201 & 2000-01-21 & Jansen & - & 39.9 & 20.7 \\ 
            0104660101 & 2000-09-17 & Watson & 22.3 & - & - \\ 
            0104660301 & 2000-11-25 & Watson & - & 20.7 & 19.6 \\ 
            0113020201 & 2001-11-19 & Aschenbach & - & 31.5 & 25.0 \\ 
            0144530101 & 2003-05-10 & McCray & - & 46.8 & 46.8 \\ 
            0406840301 & 2007-01-17 & Haberl & 53.3 & 74.4 & 76.0 \\ 
            0506220101 & 2008-01-11 & Haberl & 61.2 & 80.7 & 83.4 \\ 
            0556350101 & 2009-01-30 & Haberl & 57.3 & 79.0 & 81.7 \\ 
            0601200101 & 2009-12-11 & Haberl & 70.8 & 85.5 & 85.5 \\ 
            0650420101 & 2010-12-12 & Haberl & 46.1 & 57.9 & 60.7 \\ 
            0671080101 & 2011-12-02 & Haberl & 56.4 & 67.8 & 69.0 \\ 
            0690510101 & 2012-12-11 & Haberl & 52.9 & 38.0 & 66.7 \\ 
            0743790101 & 2014-11-29 & Haberl & 50.0 & 63.6 & 65.6 \\ 
            0763620101 & 2015-11-15 & Haberl & 51.3 & 61.3 & 62.1 \\ 
            0783250201 & 2016-11-02 & Haberl & 44.2 & 62.8 & 66.6 \\ 
            0804980201 & 2017-10-15 & Haberl & 24.4 & 48.4 & 50.8 \\ 
            0862920201 & 2020-11-24 & Haberl & 49.5 & 63.3 & 64.9 \\ \hline
            Totals & ~ & ~ & 639.7 & 921.6 & 945.1 \\ \hline
        \end{tabular}
    
\end{table}

Taking advantage of a large effective area, we also employed imaging information of the \xmm\ observations to study the energy characteristics of the diffuse emission. The images were produced with the SAS task \it evselect \rm and the exposure maps were calculated with script {\it eexpmap}. To subtract the background of the instruments, we used script \it evqpb \rm to simulate it. Then we made mosaic images of MOS2 and pn with the ESAS script {\it merge\_comb\_xmm\rm}. The out-of-time events of the bright source SN 1987A were also subtracted in this step. If necessary, the mosaic images were smoothed with the ESAS task {\it adapt\_merge\rm}.

For spectral analysis, we used commands \it mos-spectra \rm and \it pn-spectra \rm  to extract the spectra of different regions and generate related response files, while commands \it mos\_back \rm and \it pn\_back \rm are employed to estimate the quiescent particle background (QPB), which will then be subtracted from the source spectra. All the spectra are combined in accordance with the instruments (i.e., MOS1, MOS2, and pn). The HEAsoft (version 6.29) scripts \it ftmarfrmf \rm and \it ftaddrmf \rm were used to correctly weight the response matrices according to the effective areas and exposure times. The combined spectra were then binned by the task \it grppha \rm for signal-to-noise ratio greater than 7. \textit{Xspec} \citep[version 12.12.0]{xspec} with \textit{AtomDB} (version 3.0.9) was used for spectral analysis. 

\section{Results}
\label{sec:res}
\subsection{Point-like sources}
\label{sec:pls}
There are 17 point-like sources detected and analyzed in our study with \cxo\ observations as listed in Table \ref{tab:pls}, among which sources 3, 10-11, and 14-17 are newly detected, while others were listed previously in \citet{Kavanagh2019}. Source 3 is projected just outside the northern shell of the MCSNR J0536-6913. Sources 4-6 are in the center, sources 10, 16, and 17 are located outside, and other sources seem projected basically along the shell-like structure of the SB. 
As shown in the RGB image (Fig. \ref{fig:rgb}), these sources have different ``colors'', indicative of different stellar surface temperatures or different radiation mechanisms. Limited by the low count rates, we calculated their hardness ratio and roughly estimated their absorption-corrected luminosities. We also searched the SIMBAD astronomical database \citep{Wenger2000} for potential optical counterparts within 1.5$''$, or 0.36\,pc (at a distance 50\,kpc), and gave their possible physical properties in Table \ref{tab:plsop}. 

\begin{table*}
    \centering
    \setlength\tabcolsep{4pt}
    \caption{basic information of 17 point-like sources}
    \label{tab:pls}
    \renewcommand\arraystretch{1.5}
    \hspace{-20pt}    
    \begin{threeparttable}
        \begin{tabular}{ccccccccc}
        \hline
            \multirow{2}{*}{source} & \multicolumn{2}{c}{FK5 coodinates (J2000)$^a$} & Net cts$^b$ & Net flux$^b$ & significance ($\sigma$) & Hardness$^c$ & Luminosity$^d$ \\
            & RA & Dec. & (count) & ($10^{-7}\,\rm s^{-1}\,{cm}^{-2}$) & ($\sigma$) & ratio & ($10^{33}\,\rm erg\,{s}^{-1}$) \\ \hline
            1 & 05:36:33.18 & -69:11:40.88 & $216.86\pm18.10$ & $48.49\pm4.08$ & 11.98 & $0.67\pm0.17$ & 7.05 \\ 
            2 & 05:36:25.04 & -69:10:05.30 & $45.04\pm8.44$ & $11.55\pm2.16$ & 5.34 & $0.14\pm0.27$ & 2.07 \\ 
            3 & 05:36:20.41 & -69:13:02.67 & $36.36\pm6.60$ & $16.89\pm3.08$ & 5.51 & $0.45\pm0.30$ & 2.49 \\ 
            4 & 05:36:06.37 & -69:11:47.62 & $29.49\pm6.69$ & $6.28\pm1.43$ & 4.41 & $-0.52\pm0.34$ & 1.49 \\ 
            5 & 05:35:59.81 & -69:11:22.01 & $37.19\pm7.80$ & $7.70\pm1.62$ & 4.77 & $-0.55\pm0.32$ & 1.80 \\ 
            6 & 05:35:59.96 & -69:11:50.83 & $44.91\pm7.35$ & $9.18\pm1.51$ & 6.11 & $-0.46\pm0.25$ & 2.10 \\ 
            7 & 05:36:00.84 & -69:09:26.23 & $43.91\pm8.25$ & $9.75\pm1.83$ & 5.32 & $0.67\pm0.32$ & 1.43 \\ 
            8 & 05:35:57.18 & -69:09:14.58 & $387.37\pm24.98$ & $86.90\pm5.58$ & 15.51 & $0.58\pm0.11$ & 13.16 \\
            9 & 05:35:48.17 & -69:09:33.32 & $65.19\pm13.38$ & $13.79\pm2.89$ & 4.87 & $0.09\pm0.30$ & 2.49 \\ 
            10 & 05:35:43.10 & -69:09:20.24 & $60.52\pm8.50$ & $13.26\pm1.86$ & 7.12 & $0.29\pm0.21$ & 2.23 \\ 
            11 & 05:35:40.03 & -69:10:36.88 & $30.70\pm7.23$ & $6.31\pm1.49$ & 4.24 & $0.70\pm0.43$ & 0.91 \\ 
            12 & 05:35:42.40 & -69:11:51.54 & $180.31\pm14.38$ & $37.1\pm2.97$ & 12.53 & $0.64\pm0.13$ & 5.43 \\ 
            13 & 05:35:42.86 & -69:12:07.02 & $119.95\pm12.78$ & $26.40\pm2.81$ & 9.39 & $0.91\pm0.20$ & 3.46 \\ 
            14 & 05:35:42.24 & -69:12:34.11 & $21.23\pm5.62$ & $5.36\pm1.43$ & 3.77 & $0.02\pm0.38$ & 0.99 \\ 
            15 & 05:35:33.14 & -69:12:31.98 & $35.65\pm7.36$ & $8.45\pm1.74$ & 4.84 & $0.95\pm0.42$ & 1.01 \\ 
            16 & 05:35:23.32 & -69:11:01.11 & $104.95\pm11.07$ & $25.04\pm2.64$ & 9.48 & $0.65\pm0.18$ & 3.39 \\ 
            17 & 05:35:23.61 & -69:11:44.68 & $32.08\pm7.36$ & $7.64\pm1.75$ & 4.36 & $0.61\pm0.40$ & 1.05 \\ \hline
        \end{tabular}
        
        \begin{tablenotes}
            \item
            $^a$ The coordinates are given by the centroid of the ellipse regions produced with the script \it wavdetect\rm. \\
            $^b$ The energy band is 0.5--1.2\,keV. Only 5 observations directly aiming at \sbd\ were used here for the sake of a good PSF and effective area. The error is weighted assuming the counts are subject to Poisson distribution. \\
            $^c$ The hardness ratio is defined as HR=(H-S)/(H+S), where H and S are the net fluxes in 1.5-7.0\,keV and 0.5-1.5\,keV, respectively. \\ 
            $^d$ The unabsorbed luminosity is estimated by assuming the effective energy in 1.5-7.0\,keV and 0.5-1.5\,keV is 3.5\,keV and 1.1\,keV, respectively, with the LMC absorption $N_\mathrm{H}\sim7\times10^{21}\rm cm^{-2}$. The distances to the sources are approximated as 50\,kpc, the distance to the LMC; but for remote background AGNs, the distances are undoubtedly severely underestimated here.
        \end{tablenotes}
    \end{threeparttable}
\end{table*}

\begin{table}
    \setlength\tabcolsep{3pt}
    \centering
    \caption{6 point-like sources with optical counterparts}
    \label{tab:plsop}
    \begin{tabular}{ccccccc}
    \hline
        \multirow{2}{*}{source} & optical & dist. & V & B-V & spectral & likely \\
        & counterpart & ($''$) & (mag) & (mag) & type & identities \\ \hline 
        4 & Sk -69 212 & 0.07 & 12.29 & -0.15 & O6If & BSG \\ 
        5 & BAT99 79 & 0.21 & 13.58 & 0.31 & WN7ha+OB & WR \\
        6 & BAT99 80 & 0.24 & 13.24 & -0.11 & WN5h:a & WR \\
        12 & BAT99 69 & 1.21 & 17.70 & ~ & WC4 & WR \\
        13 & ST92 2-14 & 1.77 & 15.27 & 0.23 & O5.5III & AGN \\
        14 & BAT99 67 & 0.53 & 13.89 & 0.03 & WN5ha & WR \\ \hline
    \end{tabular}   
\end{table}

\begin{figure}
    \centering
    \includegraphics[width=0.48\textwidth]{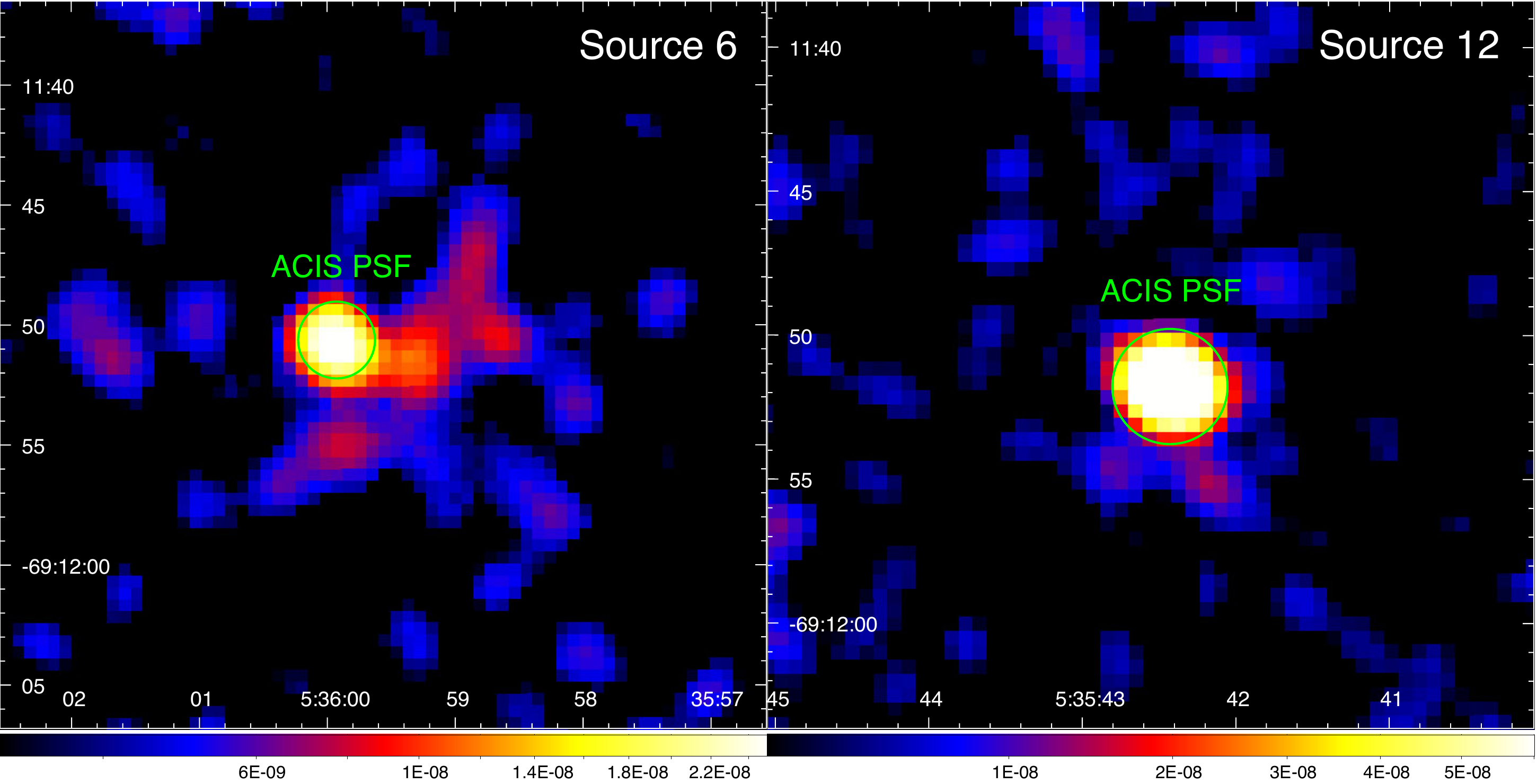}
    \caption{\cxo\ flux (unit: photon\,cm$^{-2}$\,s$^{-1}$) image of source 6 (left) and 12 (right) in 0.5-7.0\,keV band. Both images are smoothed in a scale of 3 pixels ($\sim1.5''$), and green circles mark the size of 90 per cent enclosed count fraction. Some faint clumps appear to the west of source 6 and the south of source 12, and the diffuse morphology probably arises from more than one WR star in associated sub-clusters.}
    \label{fig:point_6_12}
\end{figure}

All of the three point-like sources located in the center of \sbd, sources 4, 5, and 6, correspond to post-main-sequence bluish stars (e.g. blue super giants and Wolf-Rayet (WR) stars) seen in the optical band and have the softest spectra among the sources. The fast stellar winds of the bluish stars can produce shocks to emit the X-ray emission. Noticeably, we found that source 6 seems to have a diffuse structure extending several arcseconds as shown in Fig. \ref{fig:point_6_12}, exceeding the size of the PSF. Thus, source 6 is probably not a single point-like source but a group of sources jammed in a small space that cannot be resolved clearly. Actually, BAT99 80, the potential optical counterpart of source 6, is not only a binary candidate itself but also a member of the compact sub-association HD 269828 with nearly 20 massive stars including 3 WR stars within 10$''$ \citep{Testor1993}. Unfortunately, the limit of angular resolution of the X-ray detectors and the low count rates make further spectral analysis difficult.

Sources 12, 13, and 14 are the other three sources with potential optical counterparts within 1.5$''$. Source 12 was considered as the WR star Brey~58 \citep{Bamba2004, Lopez2020}. Here we come up with an alternative possibility that this source may be associated with one of the two other stars, the long-time variable OGLE LMC-LPV-72734 and the WR star Brey~58a, for smaller projected angular distances ($1.03''$ and $1.17''$, respectively, according to SIMBAD \citep{Wenger2000}) than Brey~58 ($2.79''$). Like source 6, the emission of source 12 may also arise from stars and their interaction in a compact sub-association containing several post-main-sequence massive stars \citep{Testor1993}. Also, Fig. \ref{fig:point_6_12} shows that source 12 probably consists of more than one point-like source. Source 13 has a quite hard spectrum with little counts below 1.5\,keV, suggesting it being fiercely absorbed, so we hold the same opinion as \citet{Bamba2004} and \citet{Lopez2020} that it is an AGN rather than an O-type star. Source 14 has never been detected and studied in previous studies and its relatively soft emission may be indicative of the X-ray counterpart of the WR star Brey~56.

Among the point-like sources associated with bluish stars, sources 12 and 14 have significantly harder spectra than sources 4, 5, and 6 located in the SB center as indicated by the hardness ratios. The spectral difference in the foreground absorption (see Section \ref{sec:spec}), their environment due to distance to the bubble center (see Section \ref{sec:mo}), and radiation mechanisms \citep[e.g.][]{Zhekov2007} may lead to this phenomenon. 

Among the sources without optical counterparts, source 2 and source 9 are relatively soft in spectra, which may be indicative of low absorption or low temperature. It is not clear whether they are independent sources or knot structures on the shell of \sbd. On the contrary, sources 1, 3, 7, 8, 10, 11, and 15--17 have hard spectra. Thus, we suppose that they are likely to be background AGNs. Stellar remnant is another possibility and \citet{Kavanagh2015} and \citet{Lopez2020} raised this notion when discussing the identity of source 1 and source 3. 

\subsection{Diffuse emission}
\label{sec:dif}

\subsubsection{Spatial distribution of hardness ratio}
\begin{figure}
    \centering
    \includegraphics[width=0.48\textwidth]{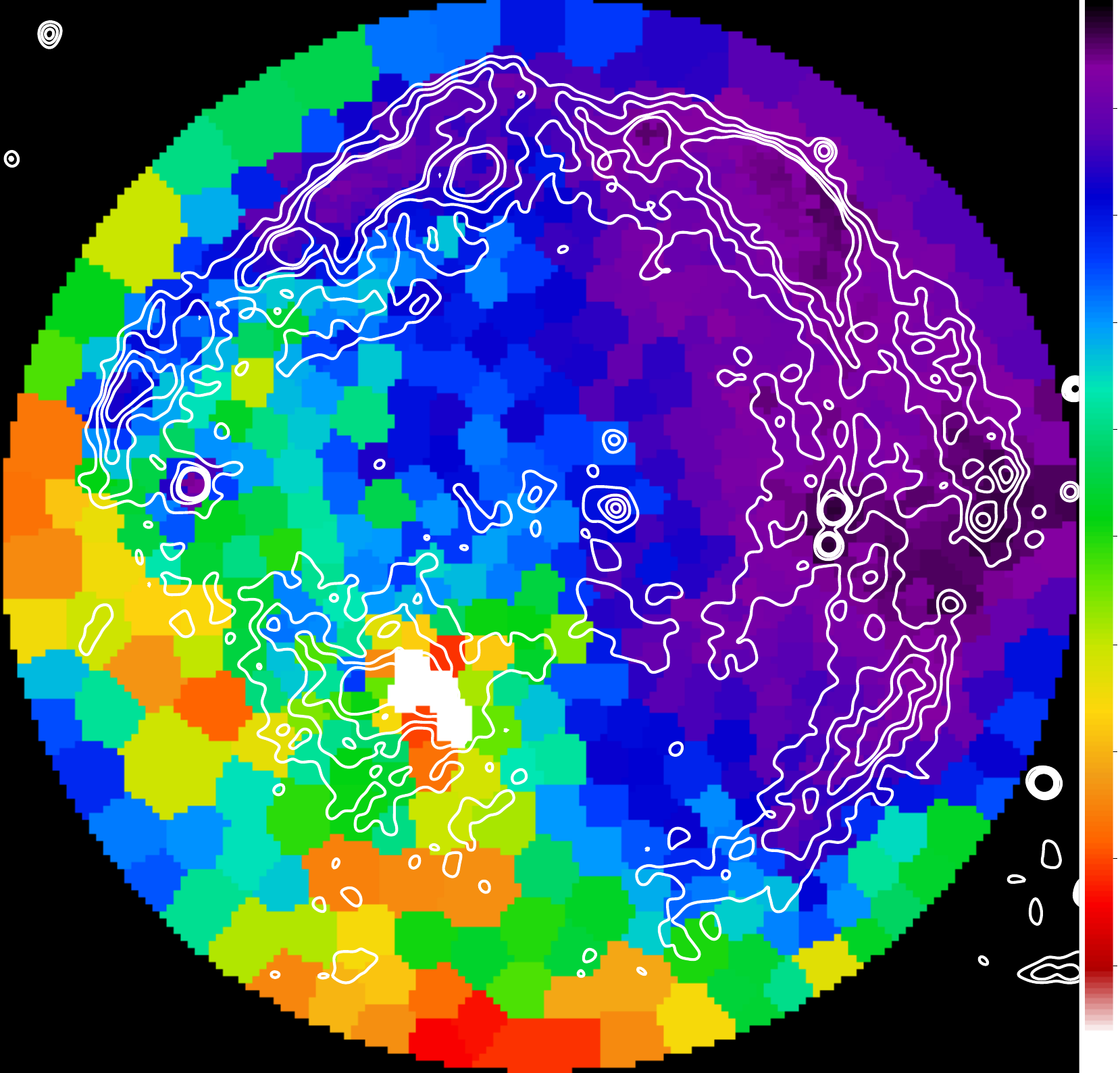}
    \caption{The hardness ratio map of \sbd\ produced with the \xmm\ MOS2 and pn observations with the QPB background subtracted. A hard band 2.0--5.0\,keV and a soft band 0.4--1.5\,keV  are used to produce the ratio map. The map is binned for an average signal-to-noise ratio of 10 and the hardness is shown to increase from red to purple. The contours are made with the high-angular-resolution \cxo\ observations for positional reference.}
    \label{fig:hr}
\end{figure}

Fig. \ref{fig:rgb} shows differences in photon energies between the western and the eastern parts of SB \sbd. To outline this difference, we constructed a ``hardness ratio map'' (Fig. \ref{fig:hr}) to map the flux ratio between the hard (2.0--5.0\,keV) and soft (0.4--1.5\,keV) bands
(note the difference of the definition of the hardness ratio here for the diffuse emission from that for the point-like sources given in the caption of Table\,\ref{fig:hr}). 
Assuming the thermal emission to be dominant below 1.5\,keV while the non-thermal emission is dominant above 2.0\,keV (also see the spectra in Appendix \ref{ap:spec}), the hardness ratio can reflect the relative intensity of the non-thermal emission to the thermal one. To guarantee the significance of the signals, we applied Weighted Voronoi Tessellation \citep{Cappellari2003, Diehl2006}, an adaptive binning method, to ensure an average signal-to-noise ratio of 10 per bin. The X-ray emission in the south-east of \sbd\ is significantly softer than that in the west, where the hardness ratio gets lowest at the shell of the MCSNR J0536-6913. The emission is quite hard in the west and the hardness ratio looks uniform compared to the surface brightness among different structures. The high hardness ratio and its uniform distribution in the west indicate that the soft thermal emission is either intrinsically very weak or heavily absorbed by the foreground intervening gas. The latter possibility seems favored by the presence of molecular clouds (MCs) mostly in the west of the SB \citep{Sano2017}.

\subsubsection{Spectroscopy}
\label{sec:spec}
\citet{Sasaki2022} claimed to discover a $\sim0.6$\,keV thermal emission in \sbd~West with eROSITA data, and \citet{Lopez2020} reported detection of a $\sim0.8$\,keV plasma in the south-western part of the bubble shell. However, both results are arguable for the lack of a local background taken into consideration so that the foreground and/or background emission may not be correctly subtracted. In fact, studies with more detailed spectral analysis based on the \xmm\ data \citep[e.g.][]{Kavanagh2015, Babazaki2018} had never detected thermal emission in \sbd\ West. With archival \xmm\ data of long accumulation over up to 20 years, it is reasonable to review the spatially resolved spectroscopy with a higher signal-to-noise ratio to search for signatures of thermal emission and to constrain the physical parameters.

\begin{figure}
    \centering
    \includegraphics[width=0.48\textwidth]{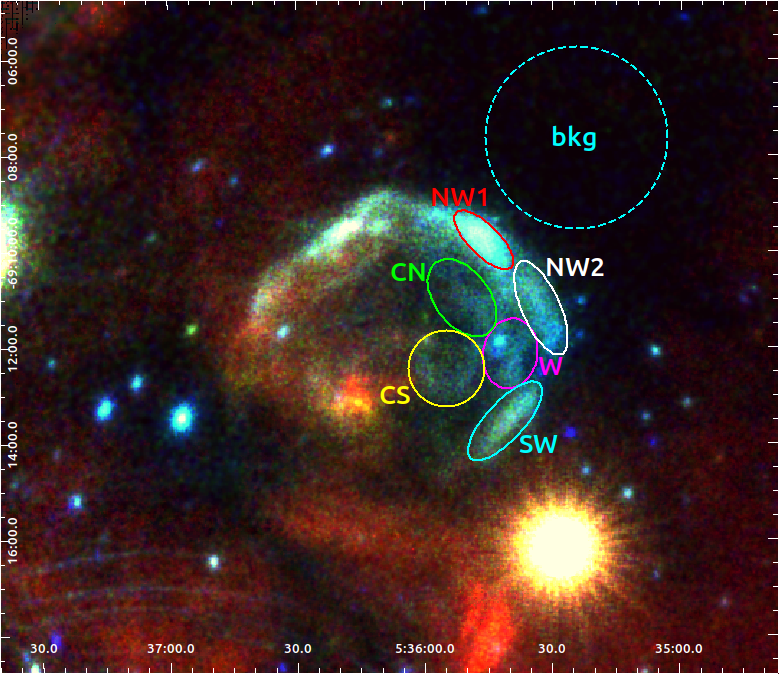}
    \caption{Regions we selected for spectral analysis. The RGB image is an \xmm\ pn and MOS2 mosaic in 0.4--1.2\,keV ({\it red}), 1.2--2.0\,keV ({\it green}), and 2.0--5.0\,keV ({\it blue}). This image reflects the difference in the emission between the west and east as shown by \citet{Kavanagh2015}. As all of the selected regions are located in the \sbd\ west, we extracted a large area to the north-west of \sbd\ as the background.}
    \label{fig:reg}
\end{figure}

The selection of regions for spectral analysis is mainly based on the morphology of SB \sbd, involving structures with high count rates to display line features in each spectrum in spite of domination by the non-thermal continuum. The regions we selected (as labeled in Fig. \ref{fig:reg}), similar to those in \citet{Kavanagh2015}, are described in detail as follows. The south-western segment of the bubble shell is selected and denoted as region SW. 
For the north-western part of the shell, the bright segment is denoted as NW1, and a fainter segment to the south-west of NW1 is denoted as NW2. \citet{Kavanagh2015} noticed diffuse emission in the interior of the SB but failed to detect thermal signals. Here we divided it into three parts. The diffuse structure just inside the western shell is labeled as region W, which is to the west of two bright point-like sources 12 and 13. On the other side of the two point-like sources, diffuse emission looks like an incomplete elliptical ring (CS+CN). 

It is noticeable that the diffuse X-ray background of the SB differs between the eastern and the western regions \citep[Fig. \ref{fig:hr}; also see][]{Kavanagh2015}, and the bright X-ray source SN\,1987A is located to the south-west of the SB. Thus, we selected a region to the north-west of the SB as the background (see Fig. \ref{fig:reg}), similar to the selection in the previous studies \citep{Kavanagh2015, Babazaki2018}, for our spectral analysis of the western region. As shown in Fig. \ref{fig:hr}, the diffuse background in almost the whole western half of \sbd\ shares a similar hardness ratio. It may thus be reasonable to assume that the true background has little spatial fluctuation and our selection of the background is feasible in this sense.

To avoid the influence of bright point-like sources, we excluded them before we selected the spectra-extracting regions. The influence of the PSF of SN 1987A and the stray light from LMC X-1 is weak enough to be neglected in our study. We analyzed the spectral data in 0.4--7.0\,keV for MOS and 0.5--7.0\,keV for pn to avoid the EPIC tail in soft energies and the line forest of pn above 7.0\,keV.

We combined all of the available \xmm\ spectra in terms of different instruments (i.e., pn, MOS1, and MOS2), instead of fitting them jointly. This helps intuitively outline the weak line features from 0.8 to 1.5\,keV above the continuum, corresponding to the lines of O, Ne, and Mg (see the spectra treated in Appendix \ref{fig:CN}). As the instrument background varies significantly in different positions of the \xmm\ detector \citep{Lumb2002, Kuntz2008}, we used the QPB spectra and added extra Gaussian components representing Al (1.49\,keV, for EMOS and EPN) and Si lines (1.75\,keV for EMOS) to model the instrumental background. The astrophysical background is modeled, but not simply subtracted, to guarantee high signal-to-noise ratios in the soft band of the source spectra where thermal emission matters. For abundance, we applied the table of solar abundance according to \citet{Anders1989} and assumed 0.5 solar abundance for the LMC \citep{Russell1992}.

The astrophysical background was characterized with a typical model comprising such components \citep{Snowden2008, Kuntz2008} as the local hot bubble (represented by an {\it apec} model without absorption), the Galactic halo (represented by an {\it apec} model with an absorbing column $N_\mathrm{H}=6\times10^{20}\,\rm cm^{-2}$ according to the \citet{Dickey1990} HI maps), and the ISM of the LMC plus the unresolved background AGNs (modeled by an {\it vapec} component plus a {\it powerlaw} component with a photon index $\Gamma=1.46$ \citep{Chen1997} and an absorption attributed to the Milky Way and the LMC). We grouped the spectra to achieve a signal-to-noise ratio $\ge5$ per bin. Since the background spectra show line features near 1.0\,keV and 1.3\,keV, we set the abundances of Ne and Mg free and added a {\it Gaussian} component near 0.9\,keV as a mimic and finally the reduced chi-squared of the fitting is 1.21. This model was simply fixed and weighted according to the backscale, or rather the region's angular size, as a part of the model when fitting on-source spectra.

For the on-source spectrum of each selected region, two fitting models with different components are tested for the source spectra: a single {\it powerlaw} component or a {\it powerlaw} component plus a {\it vapec} thermal component with variable abundances, both with the absorption of the Milky Way and the LMC. A plasma component in a non-equilibrium ionization state has also been tested, but the fitted ionization parameter $\tau=n_et$ exceeding $10^{12}\rm\,s\,cm^{-3}$ indicates a plasma component in a status of collision-ionization equilibrium (CIE). As SB \sbd\ probably may have witnessed SN explosions inside, we free the abundances of intermediate-mass elements including O, Ne, Mg, and Si. If a change of an abundance can reduce the chi-square statistics evidently, we keep its value and calculate the error range. Otherwise, the abundance is fixed to that of LMC. Since the added thermal components are very weak compared with the non-thermal components, we used the F-test to assess the difference between the two models, or in other words, the significance of the thermal components. Thus, we list the P-values or rather the possibility of a null hypothesis along with the chi-squared statistics and the degrees of freedom. In this way, we suppose that all the regions we select show evidence of thermal emission more or less.

\begin{table*}
    \centering
    \setlength\tabcolsep{5pt}
    \caption{Results of spectral analysis}
    \label{tab:fitting}
    \renewcommand\arraystretch{1.5}
    \hspace{-20pt}\begin{threeparttable}
    \begin{tabular}{cccccccccccc}
        \hline
        \multirow{3}{*}{Region} & Absorption & \multicolumn{2}{c}{Non-thermal (\it{powerlaw}\rm)} & \multicolumn{6}{c}{Thermal (\it{vapec}\rm)} & \multirow{3}{*}{$\chi^2/d.o.f.$} & \multirow{3}{*}{P-value$^d$} \\ 
        & $N_\mathrm{H}^a$ & $\Gamma$ & norm$^b$ & kT & O & Ne & Mg & Si & norm$^c$ &  &  \\ 
        & ($10^{21}\rm{cm^{-2}}$) & ~ & ($10^{-5}$) & (\,keV) & ($Z/Z_{\sun}$) & ($Z/Z_{\sun}$) & ($Z/Z_{\sun}$) & ($Z/Z_{\sun}$) & ($10^{-5}{\rm cm}^{-5}$) &  & \\ \hline
        \multirow{2}{*}{NW1} & $7.17_{-0.19}^{+0.20}$ & $2.32_{-0.03}^{+0.03}$ & $9.90_{-0.27}^{+0.28}$ & ~ & ~ & ~ & ~ & ~ & ~ & 862.3/851 & \multirow{2}{*}{$3.3\times10^{-1}$} \\
        & $7.08_{-0.23}^{+0.23}$ & $2.30_{-0.04}^{+0.04}$ & $9.47_{-0.52}^{+0.76}$ & $1.4_{-1.2}^{+0.8}$ & ~ & ~ & $2.0^e$ & ~ & $0.55_{-0.53}^{+0.81}$ & 858.9/848 & \\ 
        \multirow{2}{*}{NW2} & $11.36_{-0.34}^{+0.35}$ & $2.53_{-0.04}^{+0.04}$ & $10.20_{-0.39}^{+0.41}$ & ~ & ~ & ~ & ~ & ~ & ~ & 857.1/796 & \multirow{2}{*}{$1.8\times10^{-2}$} \\ 
        & $11.37_{-0.41}^{+0.40}$ & $2.49_{-0.05}^{+0.05}$ & $9.36_{-0.79}^{+0.35}$ & $1.6_{-0.4}^{+1.1}$ & ~ & $6.3^e$ & ~ & ~ & $1.2_{-1.0}^{+1.3}$ & 846.2/793 & \\ 
        \multirow{2}{*}{W} & $9.10_{-0.39}^{+0.40}$ & $2.38_{-0.05}^{+0.05}$ & $4.71_{-0.22}^{+0.23}$ & ~ & ~ & ~ & ~ & ~ & ~ & 537.0/524 & \multirow{2}{*}{$1.0\times10^{-3}$} \\ 
        & $9.46_{-0.44}^{+0.62}$ & $2.41_{-0.05}^{+0.05}$ & $4.88_{-0.24}^{+0.29}$ & $<0.1$ & ~ & ~ & ~ & ~ & $>9.5$  & 523.0/522 & \\ 
        \multirow{2}{*}{SW} & $4.84_{-0.17}^{+0.18}$ & $2.51_{-0.03}^{+0.03}$ & $5.37_{-0.16}^{+0.16}$ & ~ & ~ & ~ & ~ & ~ & ~ & 827.0/778 & \multirow{2}{*}{$2.2\times10^{-7}$} \\ 
        & $5.57_{-0.41}^{+0.51}$ & $2.50_{-0.05}^{+0.05}$ & $5.03_{-0.27}^{+0.31}$ & $0.29_{-0.08}^{+0.09}$ & $1.5_{-1.2}^{+3.6}$ & $1.4_{-0.8}^{+2.0}$ & $2.5_{-2.0}^{+3.7}$ & $18_{-11}^{+31}$ & $1.1_{-0.7}^{+4.2}$ & 783.7/772 & \\
        \multirow{2}{*}{CN} & $7.62_{-0.36}^{+0.38}$ & $2.52_{-0.05}^{+0.05}$ & $5.39_{-0.26}^{+0.28}$ & ~ & ~ & ~ & ~ & ~ & ~ & 604.3/561 & \multirow{2}{*}{$5.3\times10^{-3}$} \\ 
        & $8.6_{-0.9}^{+1.5}$ & $2.49_{-0.07}^{+0.08}$ & $5.25_{-0.37}^{+0.51}$ & $0.48_{-0.11}^{+0.18}$ & ~ & $2.6_{-1.7}^{+21.8}$ & $1.5_{-1.1}^{+9.2}$ & ~ & $0.9_{-0.8}^{+2.2}$ & 588.6/557 & \\ 
        \multirow{2}{*}{CS} & $4.53_{-0.20}^{+0.20}$ & $2.58_{-0.04}^{+0.04}$ & $5.07_{-0.17}^{+0.18}$ & ~ & ~ & ~ & ~ & ~ & ~ & 843.5/732 & \multirow{2}{*}{$1.2\times10^{-14}$} \\
        & $4.60_{-0.35}^{+0.40}$ & $2.41_{-0.05}^{+0.05}$ & $4.16_{-0.22}^{+0.23}$ & $0.42_{-0.04}^{+0.05}$ & $6.2_{-3.4}^{+11.4}$ & $7.6_{-3.4}^{+13.4}$ & $7.6_{-3.4}^{+13.1}$ & $12.8_{-6.7}^{+25.1}$ & $0.36_{-0.22}^{+0.23}$ & 758.4/726 & \\ \hline
    \end{tabular}

    \begin{tablenotes}
    \item
        The upper and lower limits correspond to the 90\% confidence intervals of the fit parameters. \\
        The solar abundance is based on \citet{Anders1989}. \\
        $^a$\,The model is \it phabs \rm with LMC abundance. \\
        $^b$\,Unabsorbed flux at 1\,keV in unit of $\rm{counts\,keV^{-1}\,cm^{-2}\,s^{-1}}$. \\
        $^c\,\frac{10^{-14}}{4\pi D^2}\int n_en_HdV$ in a unit of $\rm cm^{-5}$, where D is the distance to LMC in this study, and $n_e$ and $n_H$ are the density of electron and proton, respectively. \\
        $^d$\,Probability of a null hypothesis given by F-test. \\
        $^e$\, These parameters improve fitting but cannot be well-constraint.
    \end{tablenotes}
    \end{threeparttable}
\end{table*}


The fitted spectra of the selected regions are presented in Appendix \ref{ap:spec} and the spectral fitting results are listed in Table \ref{tab:fitting}. In general, the significance of the thermal emission based on the P-value is anti-correlated with the foreground absorption and the intensity of the non-thermal component.

Spectral fittings for regions NW1 and NW2 located in the north-western segment of the outer shell of \sbd\ give the highest P-value ($>0.01$) among all of the six regions. This low significance of the thermal emission is not unexpected as NW1 has the highest surface brightness and the foreground absorption in NW2 is the strongest. We let the abundances of Mg and Ne be free parameters, respectively, which indeed improves the fitting but the values cannot be well constrained. Both regions have higher gas temperatures than others; but since the significance of the thermal component is relatively low, we supposed that this arises from the degeneracy of the parameters. For these two regions, even if the properties of plasma could not be sufficiently reasonable, the emission lines of Mg and Ne are undoubtedly a signature of thermal emission.

Region W is located near the western edge of SB \sbd, where the outer shell appears broken. In the hardness-ratio map (Fig. \ref{fig:hr}), this region shows the hardest emission along with region NW2. The high hardness of these two regions is consistent with the strongest foreground absorption with a column density of the highest order $\sim10^{22}\rm\,cm^{-2}$ among the selected regions. According to our fitting, a residual appears below 1\,keV for a single power-law model, and the addition of a thermal ($kT<0.1$\,keV) component may improve this with a P-value of $\sim10^{-3}$. This result could reflect a scenario where a shock hits a cold dense cloud.

Different from NW1 and NW2, the south-western segment of the outer shell SW shows a relatively significant thermal emission (P-value $\sim10^{-7}$) with probable enrichments of intermediate-mass elements. The low absorption ($\sim5\times10^{21}\rm\,cm^{-2}$) allows a spectrum with evident line features. The temperature here is relatively low compared to those in other regions, which will be discussed in detail in Section \ref{sec:mo}. Although this region is close to the bright source SN 1987A, we exclude the possibility of contamination from the filamentary pattern due to the PSF of the instrument. It is mainly because the ejecta of SN 1987A has not been shocked yet \citep{Sun2021} and thus the metal enhancement can only account for the bubble itself.

CS and CN compose an elliptical structure in \sbd\ West. We divided it into these two regions mainly based on the curvity of the structure and the count rate. The latter distinction can be well explained by the absorbing hydrogen column densities, where the CN emission is more heavily absorbed than the CS. With a weaker non-thermal emission, thermal signals are relatively strong here compared to those at the outer shell in the same position angle, i.e. NW1. With a much lighter foreground absorption, the P-value of CS is much lower than that of any other region we selected, and the high abundances compensate the residuals below 2\,keV for the single power-law model. Although we cannot exclude the degeneracy between the normalization and the abundances, this result undoubtedly reflects line features from thermal emission. Despite different absorptions, these two regions share similar electron temperatures of 0.4--0.5\,keV with elevated metal abundances.

Overall, we reveal signatures of thermal components and potential metal overabundance in the south-western regions (CS+SW) of SB \sbd, where the foreground absorption is relatively low. The plasma is in a CIE state and cooler than that reported by \citet{Lopez2020} and \citet{Sasaki2022}. The temperature and abundance are similar to that in the eastern half of the SB \citep{Kavanagh2015}.

\subsubsection{Net flux maps}
\label{sec:flux}

The spectral analysis above shows signatures of thermal emission in \sbd\ West with enriched abundances of intermediate-mass elements (including O, Ne, Mg, and Si), which is consistent with the scenario that SB is filled with hot plasma shocked by stellar winds and SNRs and these metals gradually accumulate in the processes of stellar feedback. However, the emission is somewhere even weaker than the astrophysical background. To verify this result, we generated the narrow-band net flux map to show the spatial distribution of line emissions. With the available data, we focused on the \ion{O}{viii} ($\sim$0.6\,keV), \ion{Ne}{ix} ($\sim$0.9\,keV), \ion{Ne}{x}($\sim$1.0\,keV), and \ion{Mg}{xi} ($\sim$1.3\,keV) lines. All of them are evident emission lines of the thermal component, as represented by the cyan dashed lines in the spectra shown in Appendix \ref{ap:spec}. The \ion{Si}{xiii} line is not used due to the severe contamination of the instrumental line of MOS near 1.75\,keV. 

\begin{figure}
    \centering
    \includegraphics[width=0.48\textwidth]{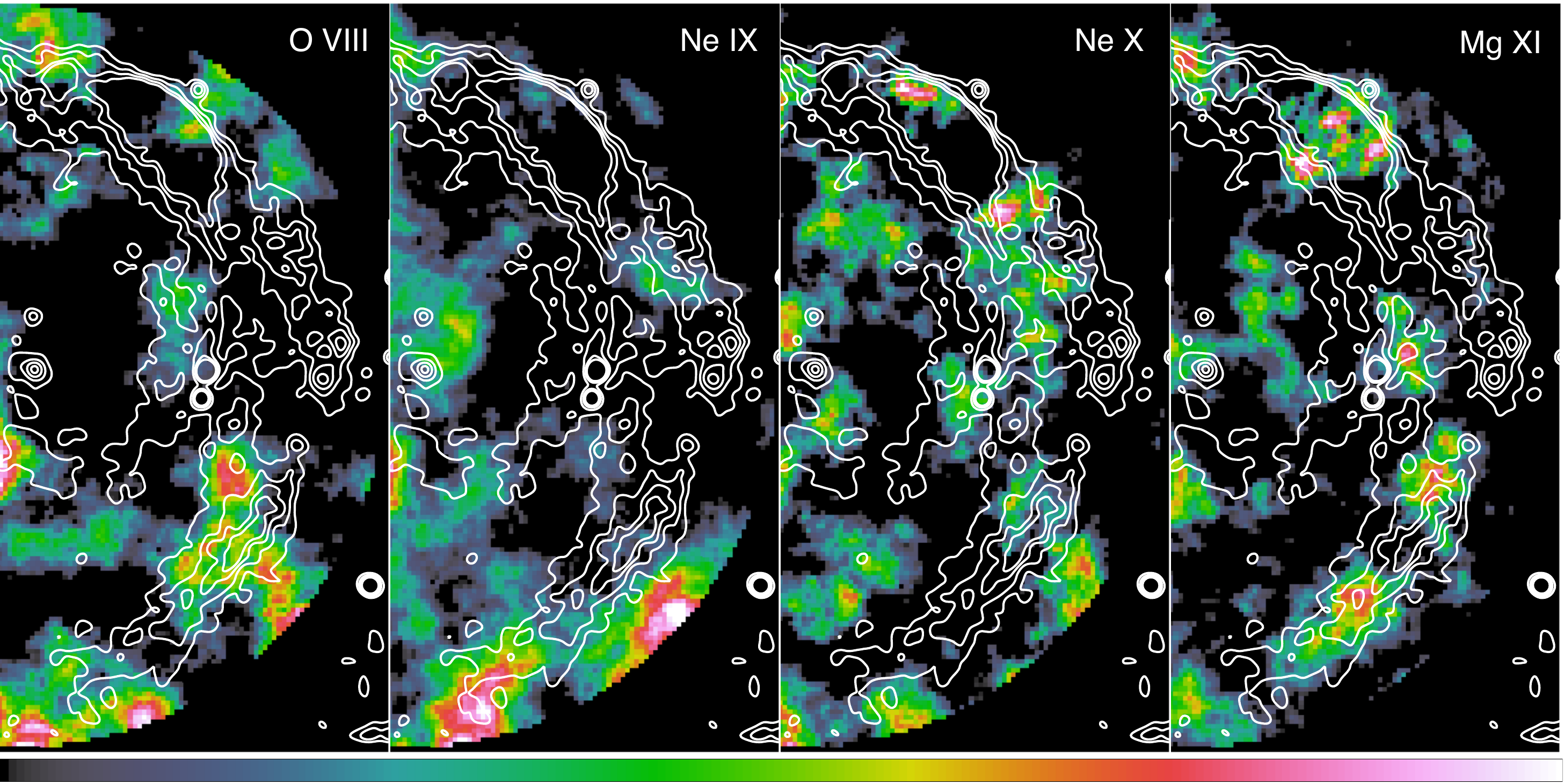}
    \caption{Net flux maps of 
    O, Ne, and Mg lines in the \sbd\ West in a linear scale. The scale ranges are 0.05--0.35 (\ion{O}{viii}), 0.20--0.70 (\ion{Ne}{ix}), 0.10--0.40 (\ion{Ne}{x}), and 0.15--1.00 (\ion{Mg}{xi}) in a unit of counts\,s$^{-1}$\,deg$^{-2}$, respectively. The contours represent the 0.5-7.0\,keV intensities in the \cxo\ observations.}
    \label{fig:narrow}
\end{figure}

As SB \sbd\ is dominated by non-thermal emission, to exhibit the thermal signals, we apply linear interpolation to subtract the underlying continuum. For the emission lines, we selected the energy intervals according to \citet{Townsley2006}, while for the continua, we selected the energy intervals on both sides whose linear interpolation is adopted for the underlying continuum. The two-side continua of \ion{Ne}{ix} and \ion{Ne}{x} lines were selected far from the line center to avoid the Fe-L forest near 0.9\,keV, and we only chose the low-energy shoulder of the \ion{Mg}{xi} lines for extrapolation to eliminate the influence of the bright Al instrumental line near 1.5\,keV. The energy intervals of the selected bands are listed in Table \ref{tab:range}. To eliminate the contamination of the QPB noise, we modeled it through the FWC data and subtracted it from each narrow-band image. After the vignetting correction, each net flux map was rebinned to a pixel size of $2.5''\times2.5''$ and smoothed with a radius of 8 pixels in a tophat convolution method, which is comparable to the size of half-energy-width of \xmm. The QPB and continuum subtracted net flux maps are shown in Fig. \ref{fig:narrow}. Equivalent width (EW) maps, another tool widely used to assess the strength of the line emission through the ratio of the net flux to the underlying continuum in the integration of line width, are not produced in our work as the non-thermal emission dominates the continuum. In such a condition, the true equivalent width, or rather the emissivity, of emission lines would be significantly underestimated. Also, an EW map, as a ratio map, requires a higher signal-to-noise ratio than a simple net counts map, which makes it difficult to map the weak thermal emission in \sbd.

\begin{table}
    \centering
    \caption{Energy ranges of bounds used for the net flux maps, where the central energy intervals are consistent to \citet{Townsley2006}.}
    \begin{threeparttable}
        \begin{tabular}{cccc}
        \hline
        Line & Center (eV) & Left (eV) & Right (eV) \\ \hline
        \ion{O}{viii} & 630--690 & 600--630 & 690--720 \\
        \ion{Ne}{ix} & 890--950 & 750--800 & 1040--1090 \\
        \ion{Ne}{x} & 990--1040 & 750--800 & 1040--1090 \\
        \ion{Mg}{xi} & 1300--1400 & 1100--1200 & 1200--1300 \\ \hline
    \end{tabular}
    \end{threeparttable}
    \label{tab:range}
\end{table}

In Fig. \ref{fig:narrow}, different lines show different spatial distributions and are not simply in proportion to the broad-band X-ray intensity. In the south of \sbd, especially in region SW, the south-western segment of the outer shell, almost all emission lines are, especially for \ion{O}{viii} and \ion{Mg}{xi}. The western edge of the SB (NW2+W) only shows weak emissions of \ion{Ne}{x} and \ion{O}{viii}, probably due to foreground absorption and/or resonant line scattering. Ne and Mg emissions are also shown in the northern part of \sbd\ West. Generally, the net flux maps are roughly consistent with the spectral analysis in Section \ref{sec:spec}, although the emission line intensities are objective to various factors including the gas temperature, density, metal abundances, and degree of ionization.

\section{Discussion}
\subsection{Properties of thermal emission}
\label{sec:prop}
Thermal emission is one of the keys to the physical nature of an SB, whose gas temperature and density reflect the evolution while the metal abundances can reveal the history of interior SN explosion. Different from temperature, the density cannot be directly measured by X-ray spectroscopy. Here, we assumed all the regions are oblate ellipsoids and roughly estimated the gas densities from the volume emission measures (VEMs) of the thermal components, which are listed in Table \ref{tab:density}.

\begin{table}
    \setlength\tabcolsep{5pt}
    \caption{Estimation of gas density in \sbd}
    \centering
    \begin{tabular}{ccccccc}
    \hline
        Region & NW1 & NW2 & W & SW & CN & CS \\ \hline
        Density & \multirow{2}{*}{$4.2_{-2.0}^{+3.1}$} & \multirow{2}{*}{$4.2_{-1.8}^{+2.3}$} & \multirow{2}{*}{$<14$} & \multirow{2}{*}{$4.1_{-1.3}^{+7.8}$} & \multirow{2}{*}{$3.5_{-1.6}^{+4.3}$} & \multirow{2}{*}{$2.2_{-0.7}^{+0.7}$} \\
        ($10^{-2}\rm\,cm^{-3}$) & & & & & & \\ \hline
    \end{tabular}
    \label{tab:density}
\end{table}

Generally, the density of thermal X-ray eminating gas in \sbd\ West is $\sim$0.01--0.1\,cm$^{-2}$, much more tenuous than typical ISM of $\sim$1\,cm$^{-3}$, although the derived densities are of great uncertainty. This indicates stellar winds and SN shocks have swept away most mass in the SB. The uncertainty may come from some factors, such as limited counts, simplistic assumptions for the gas volume, and degeneracy among the parameters. For regions NW1, NW2, and W, we suggest that the gas temperature may degenerate with the VEM where the significance of the thermal component is relatively low in Section \ref{sec:spec}. Meanwhile, as the thermal signal is much weaker than the non-thermal component, we detect them from the residual of spectral fitting with a single power-law model, which is more sensitive to line features than the continuum from the bremsstrahlung. Thus there may be degeneracy between VEM and metal abundances. For instance, region CS shows a much more significant enhancement of intermediate-mass element abundances but its density is lower compared to that in region SW. Even so, the order of magnitude of the density is still worth a reference.

Metals in an SB contain information on previous interior SNe, as they mainly originate from the ejecta of SNe rather than the stellar winds \citep[e.g.][]{Sukhbold2016}. However, it is currently difficult to make deep analysis. Firstly, the metal abundances are not constrained quite well and probably degenerate with the density as mentioned above. Secondarily, a considerable portion of the metals may be swept up into the cool dense shell of \sbd\ and the X-rays cannot trace them \citep{Weaver1977}. Thus, the total mass of metals may be seriously underestimated. Moreover, the accurate number of SNe inside the bubble is not clear yet, making it difficult to deduce the progenitors. Nevertheless, the metal enhancement in regions SW and CS is exactly the feature of recent SN(e) in \sbd. Noticeably, this is consistent with the case of some regions in the eastern half of \sbd, for which \citet{Kavanagh2015} suggest that a recent SN may have exploded near the eastern shell, accounting for the metal enrichment and radio index. The coincidence of the metal abundances, including a higher abundance of Ne and Mg than O, probably suggests that the metals uncovered in different regions share the same origin. This is reasonable as the ejecta are hardly decelerated in a low-density space until they collide with the bubble shell. The OB subclusters and massive stars are mainly distributed in the western half of \sbd\ \citep{Testor1993, Yamane2021}, where a recent SN explosion was more likely to have exploded. Also, the stellar winds from these stars may give rise to a flat radio spectrum in the west. As to the progenitors, as the age of the OB association is not more than 8\,Myr \citep{Testor1993}, we suggest that the mass of the progenitor is $\gtrsim20\,M_{\sun}$, as for stars with lower mass, the main sequence stage is longer than the age of \sbd\ \citep{Tout1996, Chen2013}. Meanwhile, \citet{Testor1993} pointed out that several stars in OB association LH\,90 are $\sim40\,M_{\sun}$ in mass, indicative of possibly an even higher mass of the supernova progenitor.

\subsection{A stellar-wind-dominant superbubble?}
\label{sec:mo}
The thermal emission of SB \sbd\ has been studied many times in the past two decades and the results differ in different studies. Without local X-ray background taken into account, \citet{Lopez2020} reported a relatively hot plasma ($\sim$0.9\,keV) in the south-west shell of \sbd, and \citet{Sasaki2022} reported a $\sim0.6$\,keV thermal component in the west of the bubble using the eROSITA data. Before that, no significant signal of thermal emission in the western part was detected \citep{Bamba2004, Yamaguchi2009, Kavanagh2015, Babazaki2018}. In our spatially resolved spectral analysis for the \sbd\ West, most of the selected X-ray bright regions with a high significance of thermal emission have a gas temperature $\sim0.4$\,keV or cooler, similar to that of the plasma in the east of \sbd\ \citep{Kavanagh2015}. We here incorporate this temperature estimate and the stellar information of the OB association LH\,90 to discuss the dynamical properties of the SB, and show that \sbd\ is mainly powered by stellar winds from the internal post-main-sequence stars as well as a recent SN explosion.

We mainly apply the model given by \citet{McCray1987} based on an adiabatic similarity solution \citep{Weaver1977} to calculate the evolution of the SB. Adopting the radius of SB \sbd\ $R=50$\,pc and applying the expansion velocity $v\sim30$\,km\,s$^{-1}$ inferred from the studies of the ambient molecular gas \citep{Sano2017, Yamane2021}, the age of the bubble is:
\begin{equation}
    \tSB=\frac{3R}{5v}=(9.8\times10^5\,\mathrm{yr})R_{50}v_{30}^{-1},
\end{equation}
where $R_{50}$ and $v_{30}$ are the bubble radius and expansion velocity in units of 50\,pc and 30\,km\,s$^{-1}$, respectively. Noticeably, this value is lower than the estimates of 4\,Myr in the previous studies \citep[e.g.][]{Smith2004} based on the age of the association \citep{Testor1993}, but is coincidentally similar to the lifespan of the WR stage, about $\sim7\times10^5\,\rm yr$ \citep{Leitherer1997}, which would probably be longer in the LMC with lower metal abundance. This probably indicates that strong stellar winds from WR stars and SN explosions may be the major energy sources of \sbd, and we will show that the mechanical luminosity of stellar winds is dominated by WR stars below. Also, this timescale allows us to suppose that the power from stellar winds can be considered constant as post-main-sequence stars we now observe just left the main-sequence stage when the bubble formed.

\citet{Testor1993} listed 8 WR stars and 25 O stars in the OB association LH\,90 which outputs mechanical energy to the SB. The mechanical luminosity of the stellar wind from a single star is estimated from the mass loss rate and the terminal velocity of the wind. For the main sequence stars, we estimate the total power of 25 O stars at $\sim5\times10^{37}\rm\,erg\,s^{-1}$ \citep[see Appendix \ref{sec:ostar} for details]{Puls1996, Kudritzki2000} but ignore that of B stars for they have much weaker wind and are not bright enough to be detected in the LMC. For the WR stars, we applied the mass loss rates of WN stars according to \citet{Hainich2014}, while those of WC stars are estimated of $\dot{M}\sim10^{-5}\rm\,M_{\sun}\,yr^{-1}$. The terminal velocities of winds from WR stars 
are adopted from \citet{Niedzielski2002} and listed along with 
and the mass loss rates in Table \ref{tab:wind}. Thus we calculate the total mechanical luminosity of stellar winds $L_w\sim1.6\times10^{38}\rm\,erg\,s^{-1}$, of which WR stars contribute $\sim1.1\times10^{38}\rm\,erg\,s^{-1}$. 

\begin{table}
    \centering
    \caption{The mechanical luminosity of WR winds}
    \label{tab:wind}
    \begin{threeparttable}
    \begin{tabular}{cccccc}
    \hline
        BAT99$^a$ & X & Sp-type$^a$ & $v_\infty^b$ & $lg\Dot{M}^c$ & $L_w^d$ \\ \hline
        67 & 14 & WN5o?+OB & 1600 & -4.94 & 0.93 \\ 
        68 & 12 & WN5--6 & 1000 & -5.46 & 0.11 \\ 
        69 & 12 & WC4 & 2600 & -5 & 2.14 \\ 
        70 & - & WC4 & 2222 & -5 & 1.57 \\ 
        77 & 6? & WN7 & 1000 & -4.87 & 0.43 \\ 
        78 & 6 & WN4 & 1600 & -4.96 & 0.89 \\ 
        79 & 5 & WN7h+OB & 1600 & -4.46 & 2.81 \\ 
        80 & 6? & O4If/WN6 & 2400 & -4.93 & 2.15 \\ \hline
        Totals & ~ & ~ & ~ & ~ & 11.03 \\ \hline
    \end{tabular}

    \begin{tablenotes}
        \item
        \textbf{Units: }$^b\rm{km\,{s}^{-1}}$; $^c\Dot{M}$: $\mathrm{M_{\sun}\,yr^{-1}}$; $^d\rm{10^{37}\,erg\,{s}^{-1}}$ \\
        \textbf{Reference: }$^a$\citet{Breysacher1999}; $^b$\citet{Niedzielski2002}; $^c$\citet{Hainich2014}(for WN)
    \end{tablenotes}
    \end{threeparttable}
\end{table}

Although \citet{McCray1987} supposed SNe would dominate the evolution of SBs after several explosions and thereafter the contribution of stellar winds is negligible, in the very early stage stellar winds are probably the main energy source. The power of stellar winds can be regarded as a minimum of the total mechanical luminosity since the overabundant intermediate-mass elements as we found in \sbd\ should be enriched by the SN explosions in the interior of the SB. Here, with the age of the bubble and the mechanical luminosity of the stellar winds, the total energy input from the winds is:
\begin{equation}
    L_w\tSB=(4.9\times10^{51}\,\mathrm{erg})R_{50}v_{30}^{-1}.
\end{equation}
This indicates that the energy from stellar winds is commensurate with explosions of $\sim$5 canonical SNe. If the efficiency of transforming such energy to cosmic rays is comparable to that of SNRs ($\sim$10\%), it could satisfy the energy requirement of $\sim5\times10^{50}$\,erg to produce the TeV emission in \sbd\ in a hadronic scenario \citep{H.E.S.S2015}.

For SNe, as their accurate number and explosion energy are still unknown to estimate the energy input, we introduce factor $A$ as the ratio of the power from SNe to that from stellar winds, i.e., the total mechanical luminosity $L=(1+A)\,L_w$. $A$ is greater than 0, and later we will constrain factor $A$ with observations.

The radius of an SB can be given by the mechanical luminosity and ambient density \citep{McCray1987}:
\begin{equation}
  \label{eq:Rw}
    R=(67\,\mathrm{pc})(L_{38}t_6^3/n_0)^\frac{1}{5},
\end{equation}
where $L_{38}=L/(10^{38}\,\mathrm{erg\,s^{-1}})$,
$t_6=\tSB/(10^6{\rm yr})$, and $n_0$ is the density of the ambient ISM in units of cm$^{-3}$. For \sbd, adopting the radius $\sim$50\,pc and the age $9.8\times10^5$\,yr calculated above, we can get the relation between the ambient gas density and the factor A:
\begin{equation}
    n_0=(6.5\mathrm{cm^{-3}})(1+A)R_{50}^{-2}v_{30}^{-3}.
\end{equation}

To verify the result above, we calculate and compare the mass of the bubble shell in two different ways. As pointed out by \citet{McCray1987}, 1/5 of total energy becomes kinetic energy of the shell, from which the mass of the shell can be estimated:
\begin{equation}
    m_s=\frac{2}{5}\frac{L\tSB}{v^2}=(1.7\times10^4\,\mathrm{M_{\sun}})n_0R_{50}^3.
\end{equation}
On the other hand, the mass interior to the SB is chiefly comprised of the ambient material swept up into the shell by the shock driven by stellar winds and/or SNe. This mass would then be
\begin{equation}
    m_s=1.4m_pn_0\times\frac{4}{3}\pi R^3=(1.8\times10^4\,\mathrm{M_{\sun}})n_0R_{50}^3,
\end{equation}
where $m_p$ is the mass of a proton. The latter estimation is model-independent, and the two estimates of mass are consistent quite well with one another.

Based on the age and mechanical luminosity of \sbd\ calculated above, we then estimate the radial profile of temperature and density in the bubble based on \citet{MacLaw1988}:
\begin{equation}
\label{eq:t}
    T(x)=(0.56\,\mathrm{keV})(1+A)^\frac{2}{7}(1-x)^\frac{2}{5}R_{50}^{-\frac{2}{7}}
\end{equation}
and
\begin{equation}
\label{eq:n}
    n(x)=(0.052\,\mathrm{cm^{-3}})(1+A)^{\frac{5}{7}}(1-x)^{-\frac{2}{5}}R_{50}^{-\frac{12}{7}}v_{30}^{-1},
\end{equation}
where $x=r/R$ is the dimensionless distance to the center of the SB. As the densities depend on the assumption of region shapes, they have much more uncertainties than temperatures which can be directly measured. Here we applied the gas temperatures of regions CN and CS, where thermal emission is the most significant and parameters are constrained relatively well. 

\begin{figure}
    \centering
    \includegraphics[width=0.49\textwidth]{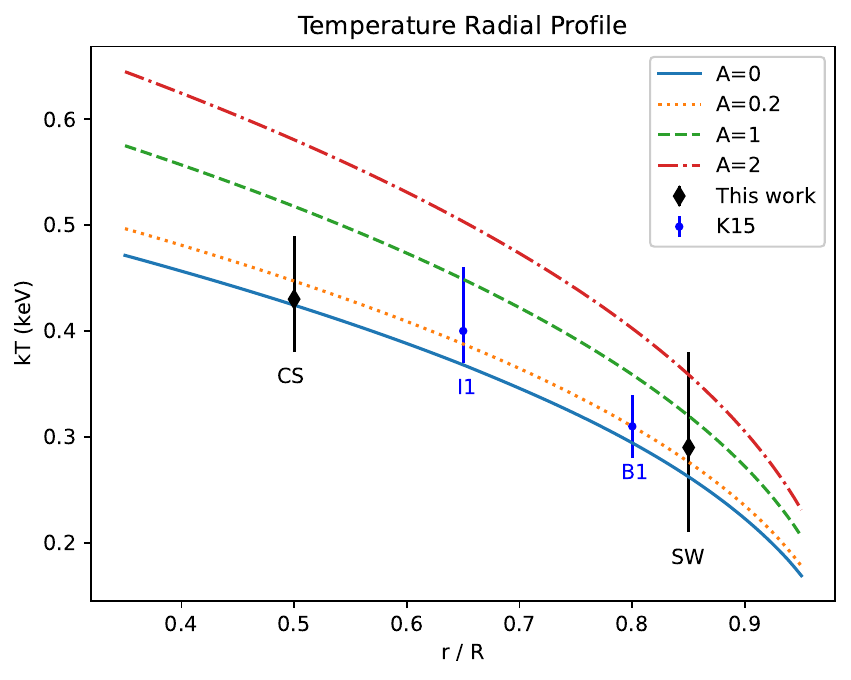}
    \caption{The radial profile of temperature given by models (lines) versus spectral results. Black diamond points are our spectral results while blue dots are inferred from \citet{Kavanagh2015}, K15 for short. Situations include pure wind (A=0, blue line), SNe comparable to the wind (A=1, green dash line), SNe dominated (A=2, red dash-dot line), and wind plus an SN energy of $\sim10^{51}$\,erg (A=0.2, orange dot line). The vertical error bars correspond to a 90\% confidence level. The medium radius is roughly estimated, and thus here we do not apply horizontal error bars.}
    \label{fig:profile}
\end{figure}

We also adopted the study of \citet{Kavanagh2015} to expand the region sample. Regions B1 and I1 therein are supposed to have thermal emission as well. We here roughly supposed the projected distance from these regions to the bubble center as the approximate distance and plot the radial profile of temperature in Fig. \ref{fig:profile}. This plot excludes scenarios where the power of SNe is equal or superior to that from stellar winds (i.e., $A\ge1$). On the other hand, the metal enhancement according to spectroscopy indicates the contribution of SNe. Thus, we suggest the energy source of \sbd\ is dominated by stellar winds and at least one SN exploded recently. For instance, $A=0.2$ corresponding to one conical SN fits the temperature radial profile quite well. Thus, the total mechanical luminosity $L=(1+A)L_w$ would amount to $2\times10^{38}$\,erg\,s$^{-1}$.

\subsection{Particle acceleration in \sbd}
Superbubbles are candidates of powerful cosmic-ray accelerators, where turbulence, multiple shocks, and shock-wind interactions could accelerate particles effectively despite a high sound velocity inside \citep{Bykov1992, Parizot2004, Vink2020, Vieu2022b}.
\sbd\ is a unique SB in the LMC for its luminous non-thermal X-ray emission up to 20\,keV \citep{Lopez2020}. Also, it is the first SB detected in TeV band \citep{H.E.S.S2015}. These all suggest particle acceleration inside it. However, either non-thermal X-ray emissions or TeV $\gamma$-ray emissions have never been accurately robustly detected in other superbubbles in the LMC including N~11 \citep{Maddox2009}, N~51D \citep{Cooper2004}, N~158 \citep{Sasaki2011}, and N~206 \citep{Kavanagh2012}, and thus one must consider the origin of particle acceleration of \sbd\ or what makes it stands out.

We suggest that the high energy input from stellar winds and exploded SNe probably account for a higher acceleration effect, which is dominated by WRs and supergiants. Compared to other SBs in the LMC as mentioned above, the number of WR stars in \sbd, which dominate the power of stellar winds, is larger than that in any other bubble in the LMC except the famous star-forming region 30\,Dor \citep{Hung2021}. The stellar winds can constantly inject energy into the bubble within $\sim$1\,Myr, much longer than the average interval between two supernova explosions in an SB $\sim3\times10^5$\,yr \citep{MacLaw1988}. Assuming the particle acceleration efficiency is $\sim$10\%, which is typical for an SNR, the total power to accelerate particles is $\sim2\times10^{37}$\,erg\,s$^{-1}$ injected by stellar winds along with an extra SN inside (see details in the subsection above). This coincides with the required energy injection of $\sim2\times10^{37}$\,erg\,s$^{-1}$ in a time-dependent model to emit the TeV $\gamma$-ray emissions in a leptonic scenario \citep{H.E.S.S2015}. In \sbd, the non-thermal X-rays, which trace the high-energy electrons, are enhanced in the north-west, coincident with the distribution of OB subclusters and massive stars \citep{Yamane2021}. This is quite similar to the case of Galactic Cygnus Superbubble with a mechanical luminosity of $\sim2\times10^{38}$\,erg\,s$^{-1}$ \citep{Aharonian2019} and evident $\gamma$-ray emissions from GeV to PeV \citep{Ackermann2011, ARGO2014, HAWC2021, LHAASO2023}. 

Another possible particle accelerator could be a recent SN in \sbd\ as we pointed out above, although it may have a low possibility \citep{Smith2004}. An SN of $10^4$\,yr ago, for instance, can raise the average mechanical luminosity of one order of magnitude larger than that of stellar winds in a short time (but in a long time scale, the energy of \sbd\ is dominated by stellar winds as discussed above). Without constant energy input, an SNR can emit non-thermal keV X-ray emission in a short timescale in which the SNR shock propagates at a velocity not lower than $\sim2-3\times10^{3}$\,km\,s$^{-1}$ \citep[e.g.][]{Vink2012, Ge2021}. Assuming a canonical SNR in the Sedov-Taylor phase \citep{Sedov1959, Taylor1950}, we estimate the timescale of emitting synchrotron X-rays as:
\begin{equation}
    t_{\rm keV}\sim(1.4\,\mathrm{kyr})\left(\frac{E_{\rm SN}}{10^{51}\,\mathrm{erg}}\right)^{\frac{1}{3}}\left(\frac{n_0}{0.04\mathrm{\,cm^{-3}}}\right)^{-\frac{1}{3}}\left(\frac{v_s}{3000\,\mathrm{km\,s^{-1}}}\right)^{-\frac{5}{3}},
\end{equation}
where $E_{\rm SN}$ is the explosion energy, $n_0$ is the gas density in the SB, and $v_s$ is the shock velocity. Compared to the time interval between two SNe of $\sim3\times10^5$\,yr \citep{MacLaw1988}, it is quite short, indicative of a very low likelihood for the presence of non-thermal keV X-rays.

Even if an SN recently exploded, stellar winds can still play a key role in particle acceleration. Compared to other SBs lacking strong stellar winds, \sbd\ allows multiple shocks \citep{Bykov1992, Parizot2004} or fierce turbulence arising from the interaction between the SNR shock and the strong winds \citep{Vieu2022} where the particles can be accelerated to higher energy up to $\sim40$\,PeV \citep{Bykov2015}. These mechanisms may dominate the cosmic ray acceleration in the SB, whereas the high sound velocity in the hot medium leads to a low Mach number of explosive motion in it. Meanwhile, the stellar winds may provide power to reaccelerate these high-energy particles and thus allow the X-ray synchrotron to sustain for a much longer time.

In an SB, the most massive stars only exist in the first several million years due to their short lifespan. Later, the stellar winds are negligible compared to supernova explosions \citep{Vieu2022b}. Without strong stellar winds, particles would not be able to be accelerated to non-thermal X-ray emitting energies a couple of thousand years after an SN explosion, probably explaining the lack of X-ray synchrotron emissions in other SBs. At an age of $\sim$1\,Myr, \sbd\ is in a young stage when the stellar winds are strong enough and some of the most massive stars begin to explode, probably indicative of efficient and constant particle acceleration. The timescale of such a stage \citep[$\sim1$\,Myr,][]{Vieu2022b} is much shorter than the lifespan of SBs \citep[$\gtrsim10^7$\,yr,][]{MacLaw1988}, corresponding to small samples of SBs emitting non-thermal X-rays like \sbd.

\section{Summary}
\label{sec:sum}
We employ deep \cxo\ and \xmm\ observations to perform spatially resolved spectroscopic X-ray studies of the western half of SB \sbd. We enlarge the sample of point-like sources in the SB interior and find large-scale signatures of diffuse thermal X-rays in the western half among the dominant non-thermal emissions. We summarize our results as follows.

1. We listed the basic information of 17 point-like sources, calculating their ratios and estimating their luminosity. We also searched for their potential optical counterparts and gave the likely associations of some of them according to hardness ratios and spectral properties. Several sources correspond to known WR stars or OB subclusters.
    
2. We applied spectral analysis and continuum-and-background-subtracted narrow-band maps to search for thermal emission in the western half of SB \sbd. Thermal signals are significant in the south-west of the bubble with a lower column density. For other regions, metal line emissions are also detected although some physical parameters cannot be constrained well. The electron temperature of the plasma is $\sim0.3-0.4$\,keV, and abundances of intermediate-mass metals including O, Ne, Mg, and Si are elevated, which may come from a recent SN.

3. We calculated the mechanical luminosity of massive stars $\sim1.6\times10^{38}\,\rm erg\,s^{-1}$, most of which is contributed by internal post-main-sequence stars, especially 8 WR stars and 3 supergiants. We suggest that SB \sbd\ is dynamically dominated by stellar winds, although it may have witnessed an SN explosion event recently, and can be blown out within $\sim1$\,Myr. The strong stellar winds, along with a recent SN, probably play a significant role in terms of its unique non-thermal X-ray emissions and TeV $\gamma$-ray emissions among the SBs in the LMC.

\section*{Acknowledgements}
This study is based on observations obtained with \xmm, an ESA science mission with instruments and contributions directly funded by ESA Member States and NASA. The research has made use of data obtained from the \cxo\ Data Archive and the \cxo\ Source Catalog, and software provided by the \cxo\ X-ray Center (CXC) in the application packages \texttt{CIAO}. Y-HC thanks Xiao Zhang for the discussion on $\gamma$-ray emissions of \sbd. This work is supported by the NSFC under grants 123B1021, 12173018, 12121003, and 12273010. LS acknowledges the support from Jiangsu Funding Program for Excellent Postdoctoral Talent (2023ZB252).

\section*{Data Availability}
The data used in this work is available via \cxo\ Data Archive (https://cda.harvard.edu/chaser) and \xmm\ Data Archive (https://www.cosmos.esa.int/web/xmm-newton/xsa).

\bibliographystyle{mnras}
\bibliography{reference}

\appendix

\section{spectral fit figures}
\label{ap:spec}
\begin{figure*}
    \centering
    \subfigure{\includegraphics[width=0.45\textwidth]{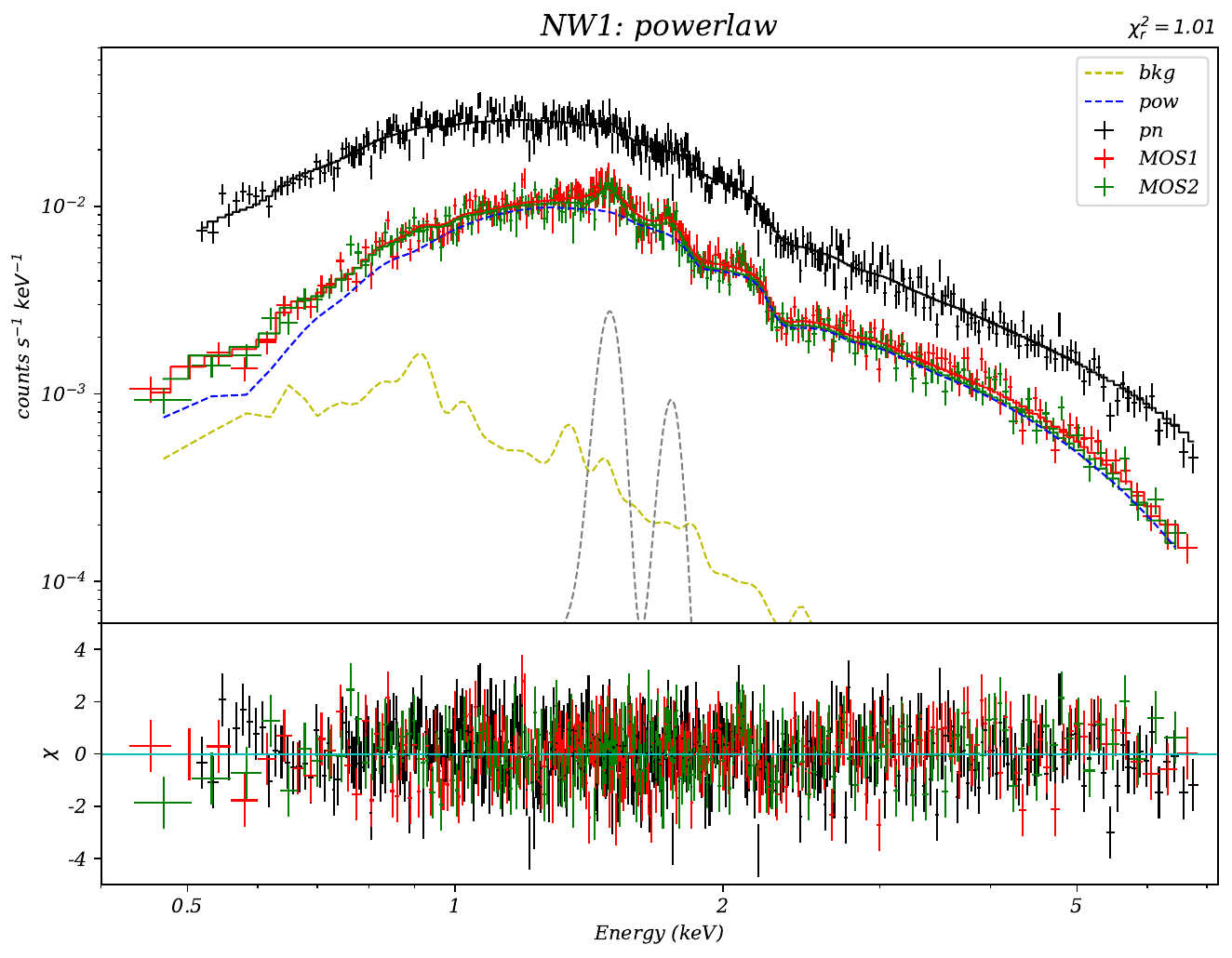}}
    \subfigure{\includegraphics[width=0.45\textwidth]{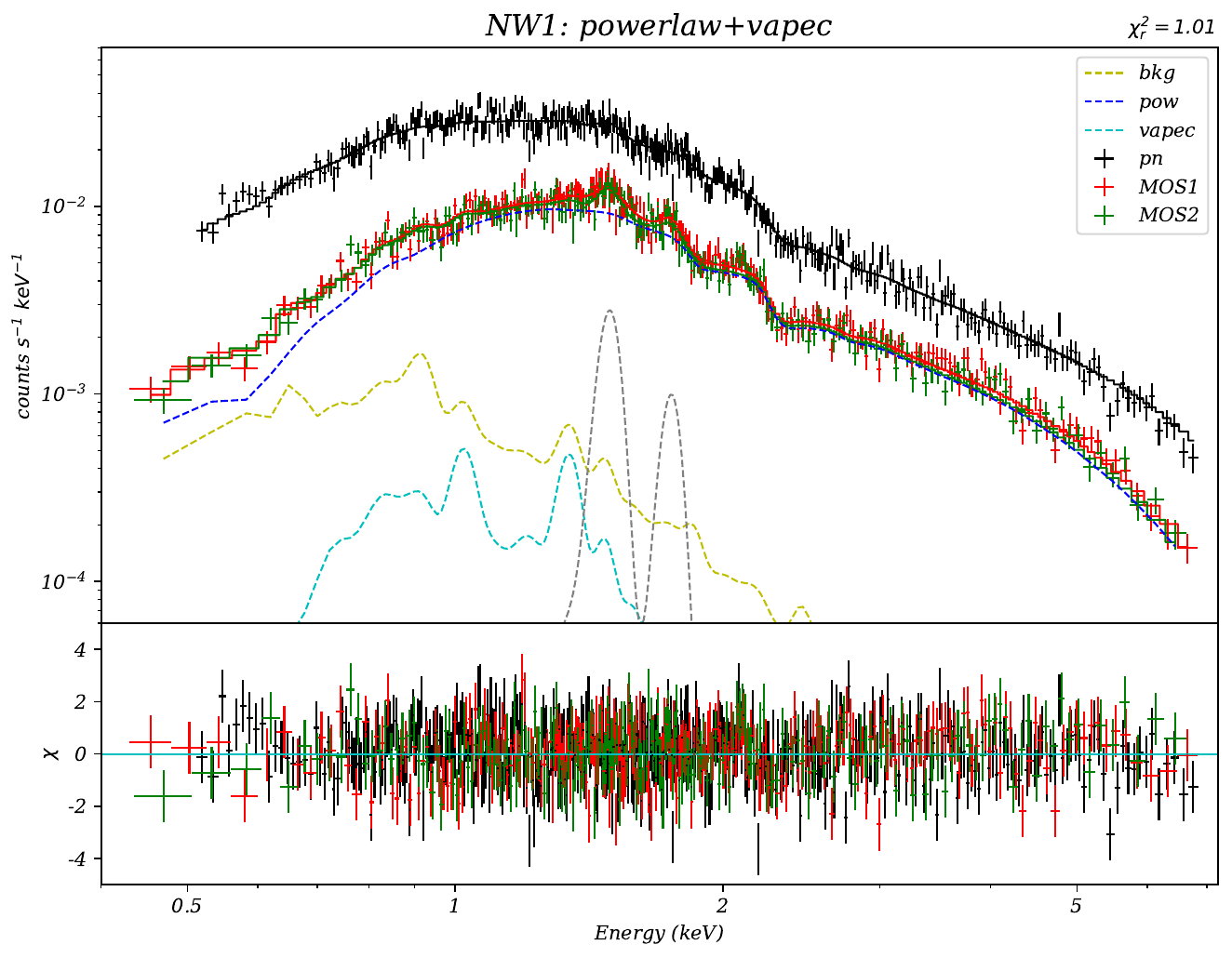}}
    \label{fig:NW1}
    \caption{The spectra and the folded models of region NW1.}
\end{figure*}

\begin{figure*}
    \centering
    \subfigure{\includegraphics[width=0.45\textwidth]{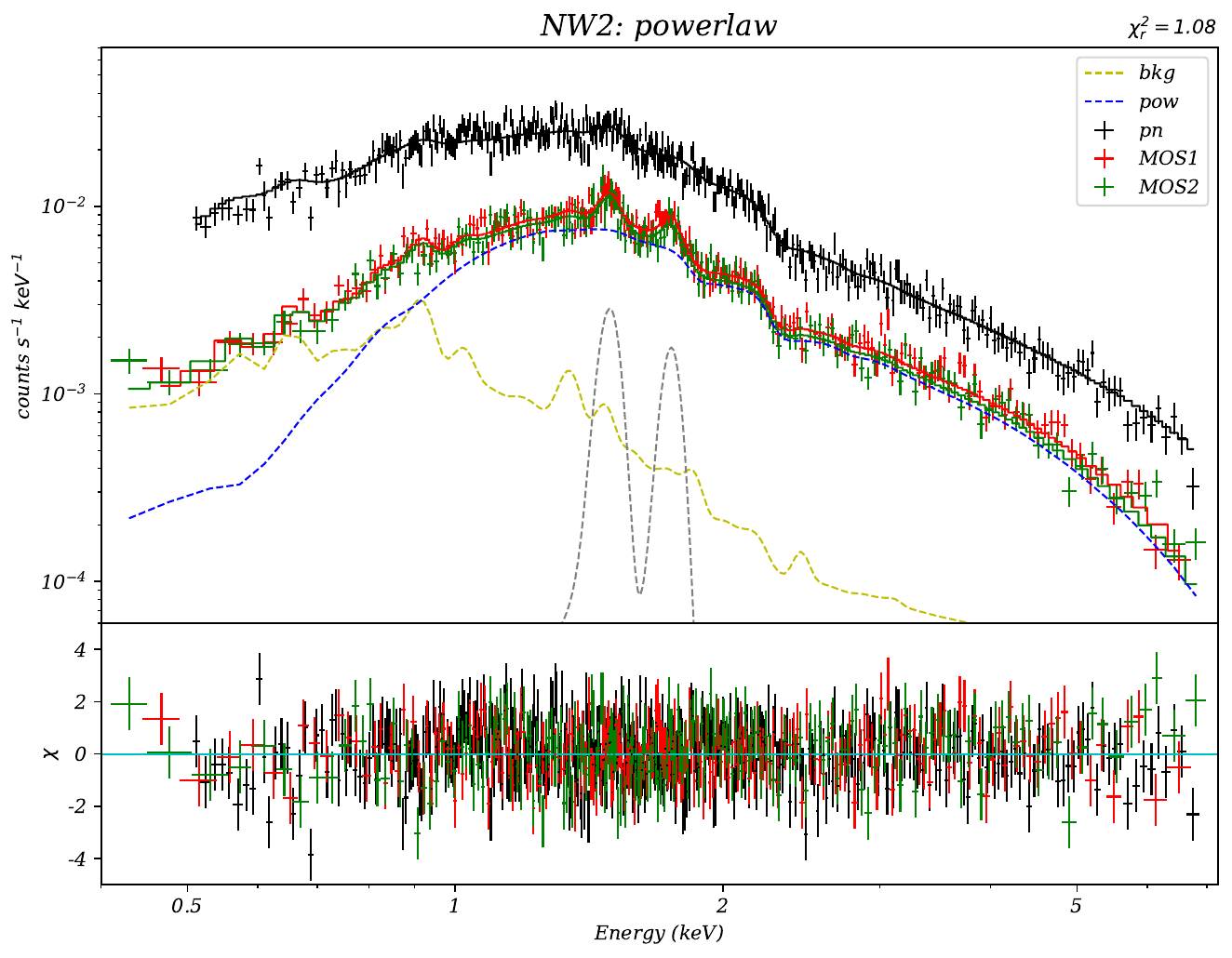}}
    \subfigure{\includegraphics[width=0.45\textwidth]{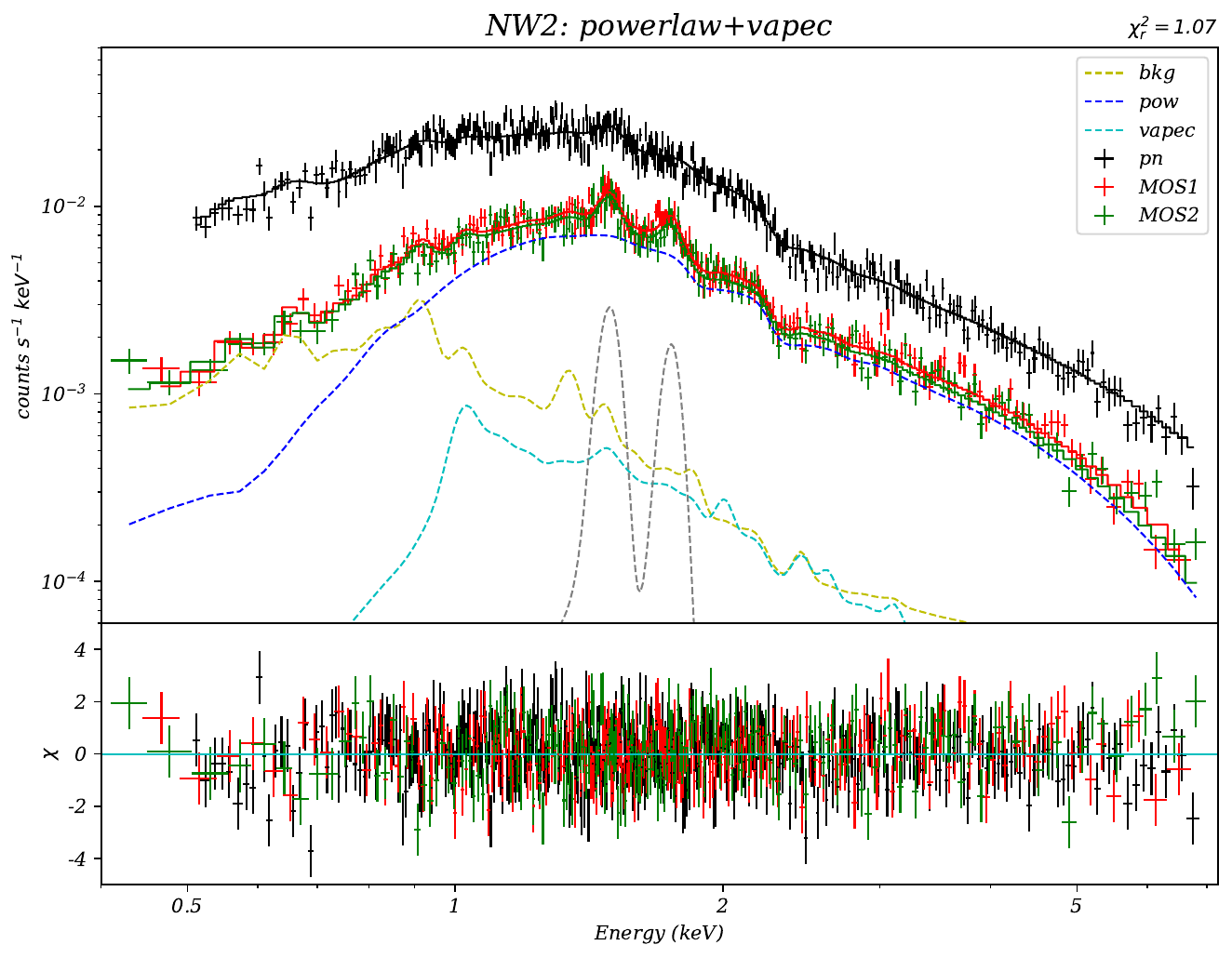}}
    \label{fig:NW2}
    \caption{The spectra and the folded models of region NW2.}
\end{figure*}

\begin{figure*}
    \centering
    \subfigure{\includegraphics[width=0.45\textwidth]{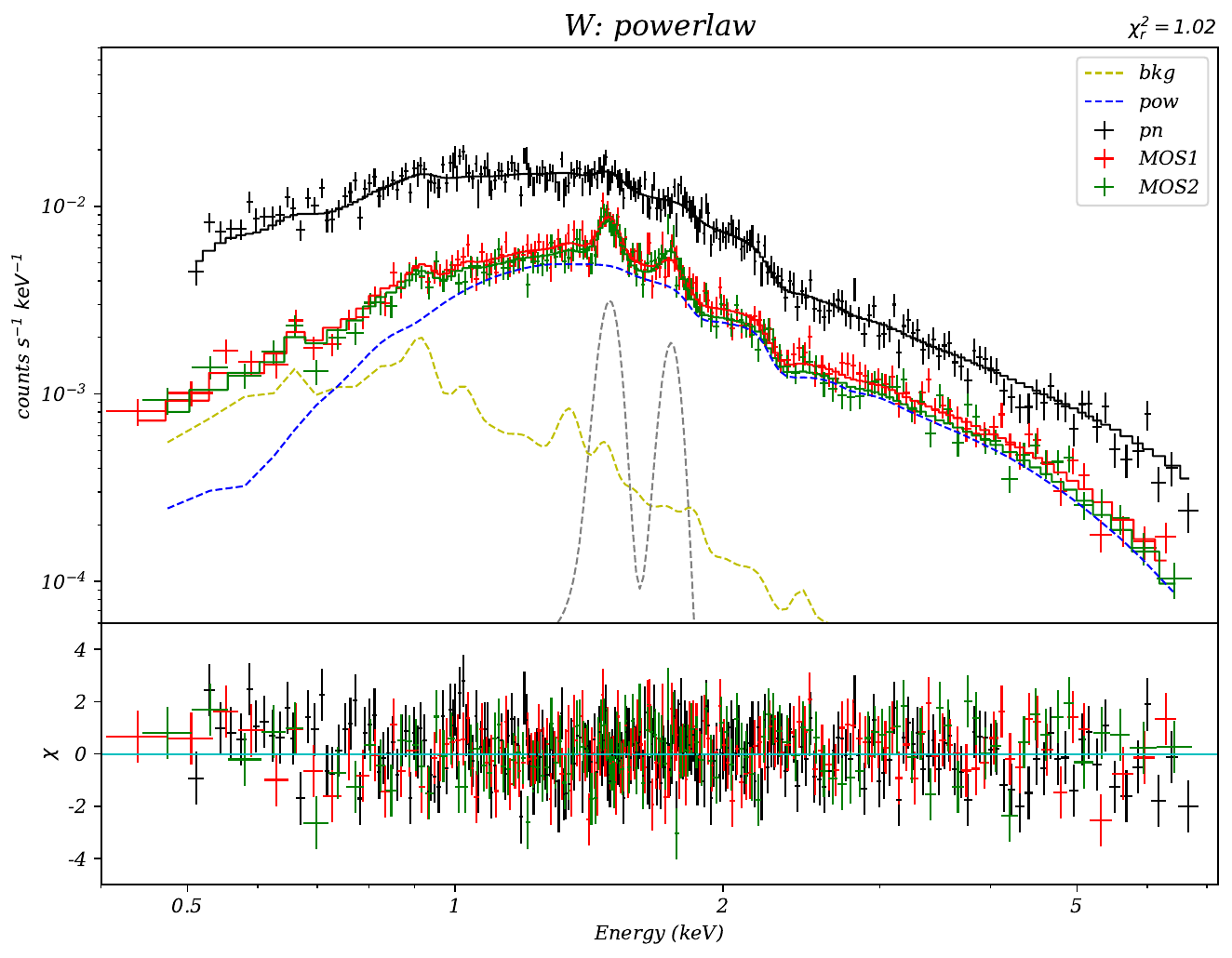}}
    \subfigure{\includegraphics[width=0.45\textwidth]{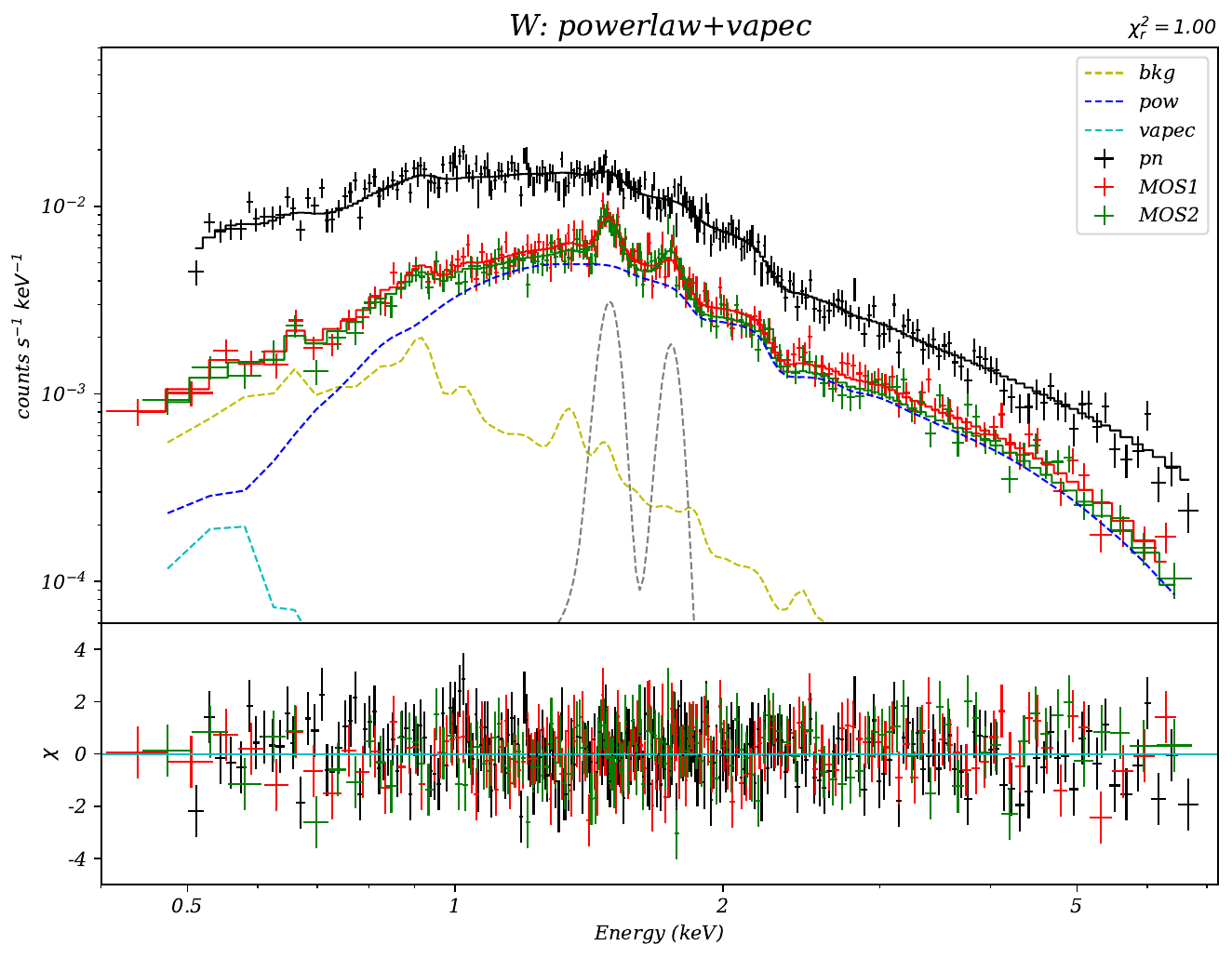}}
    \label{fig:W}
    \caption{The spectra and the folded models of region W.}
\end{figure*}

\begin{figure*}
    \centering
    \subfigure{\includegraphics[width=0.45\textwidth]{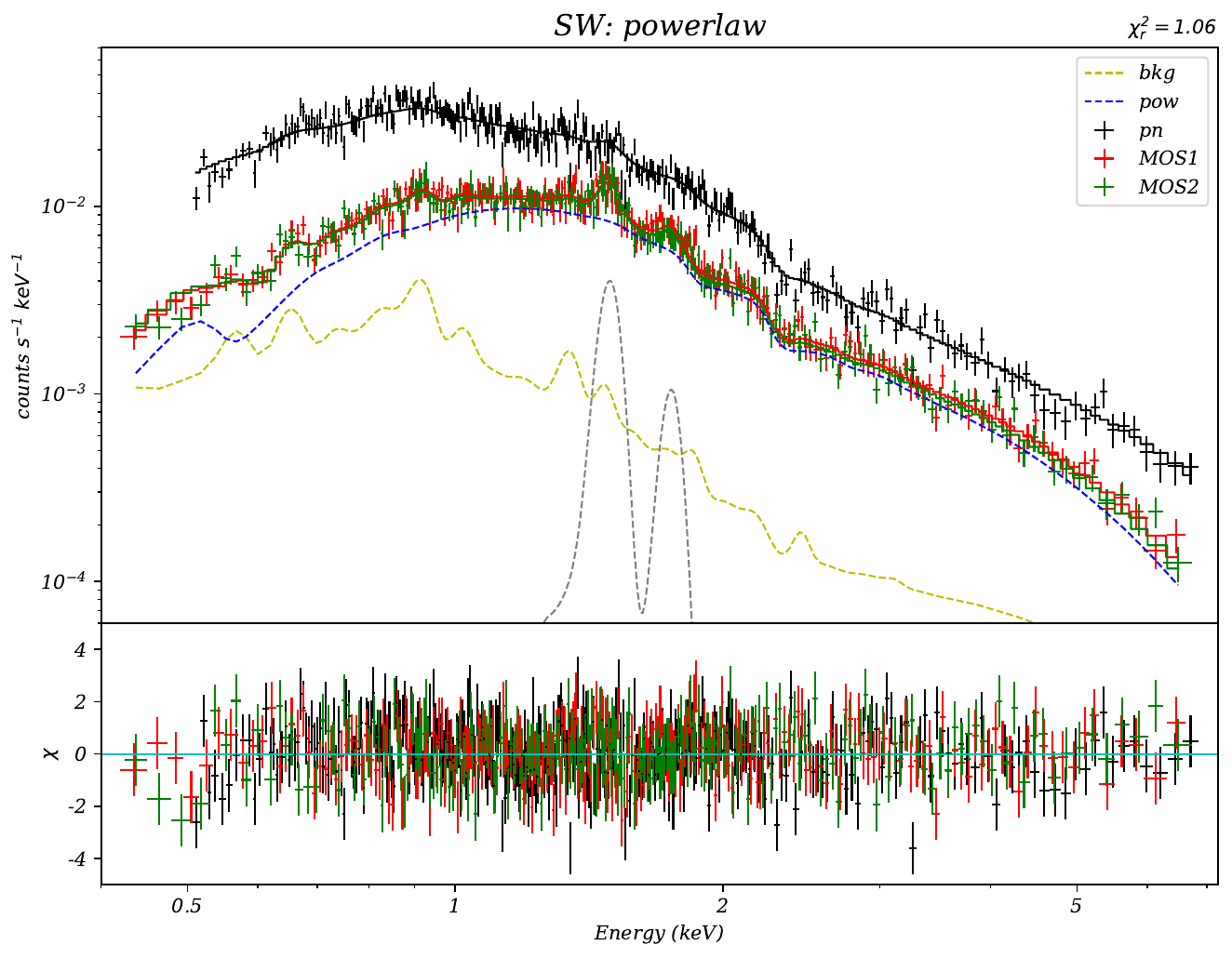}}
    \subfigure{\includegraphics[width=0.45\textwidth]{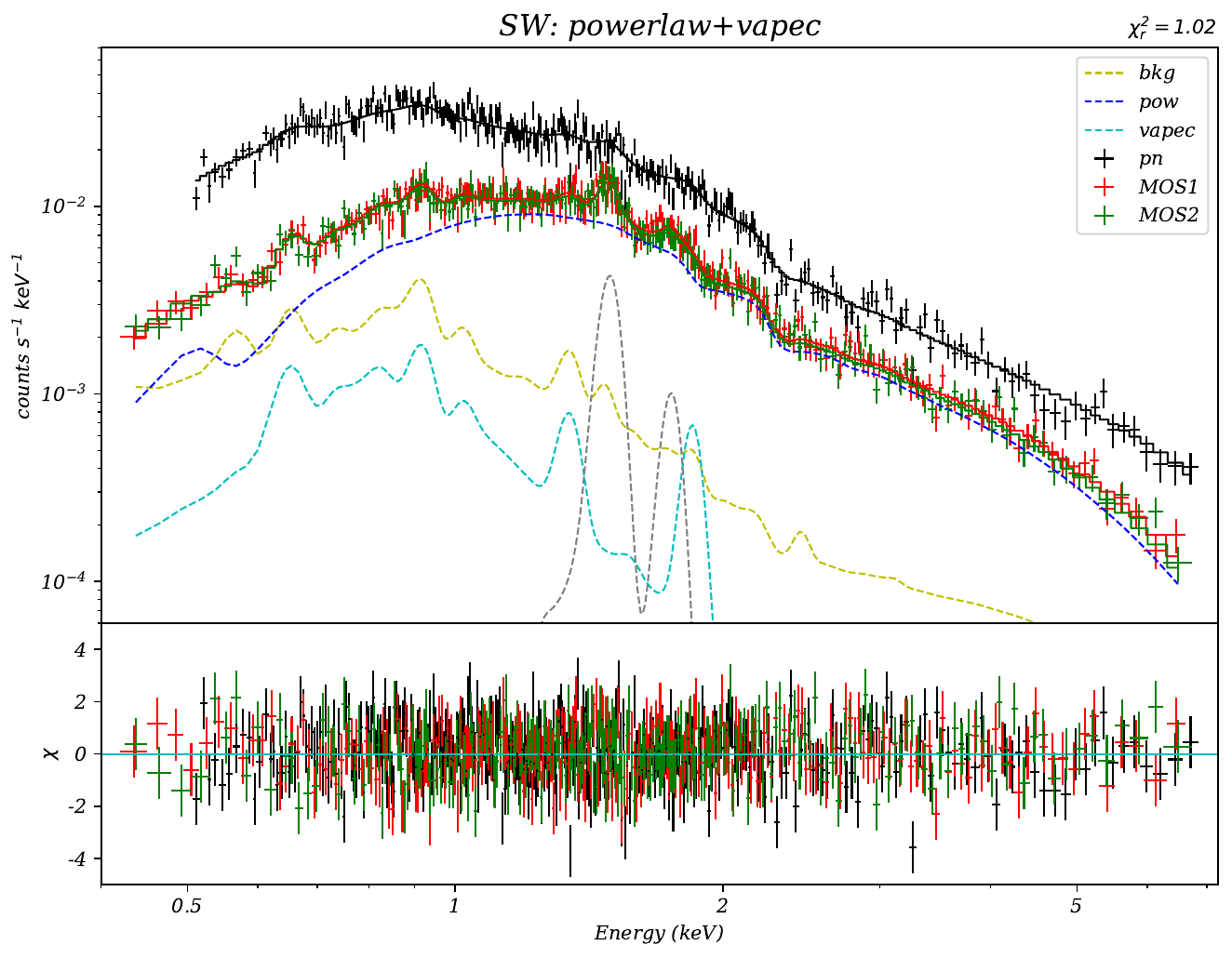}}
    \label{fig:SW}
    \caption{The spectra and the folded models of region SW.}
\end{figure*}

\begin{figure*}
    \centering
    \subfigure{\includegraphics[width=0.45\textwidth]{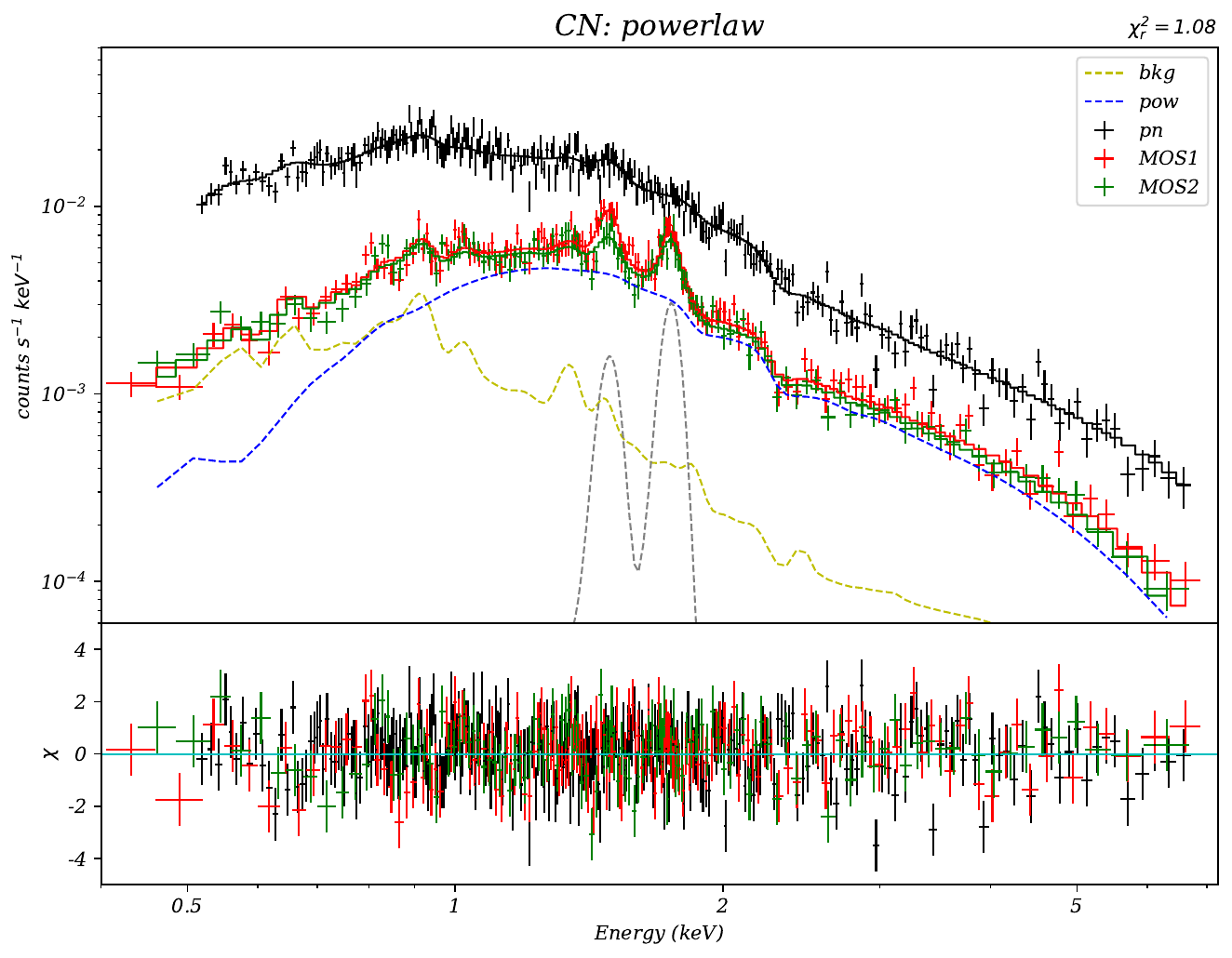}}
    \subfigure{\includegraphics[width=0.45\textwidth]{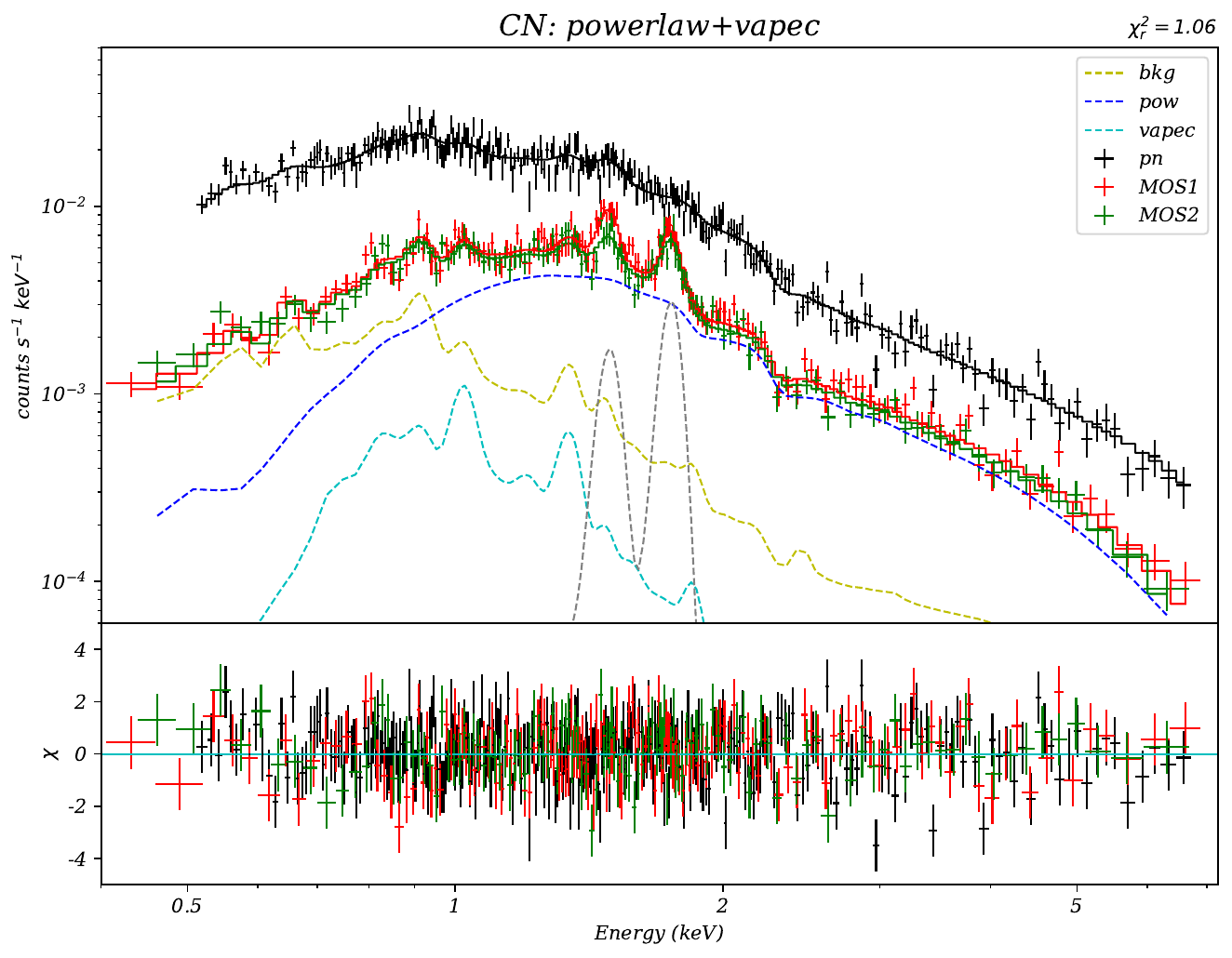}}
    \label{fig:CN}
    \caption{The spectra and the folded models of region CN.}
\end{figure*}

\begin{figure*}
    \centering
    \subfigure{\includegraphics[width=0.45\textwidth]{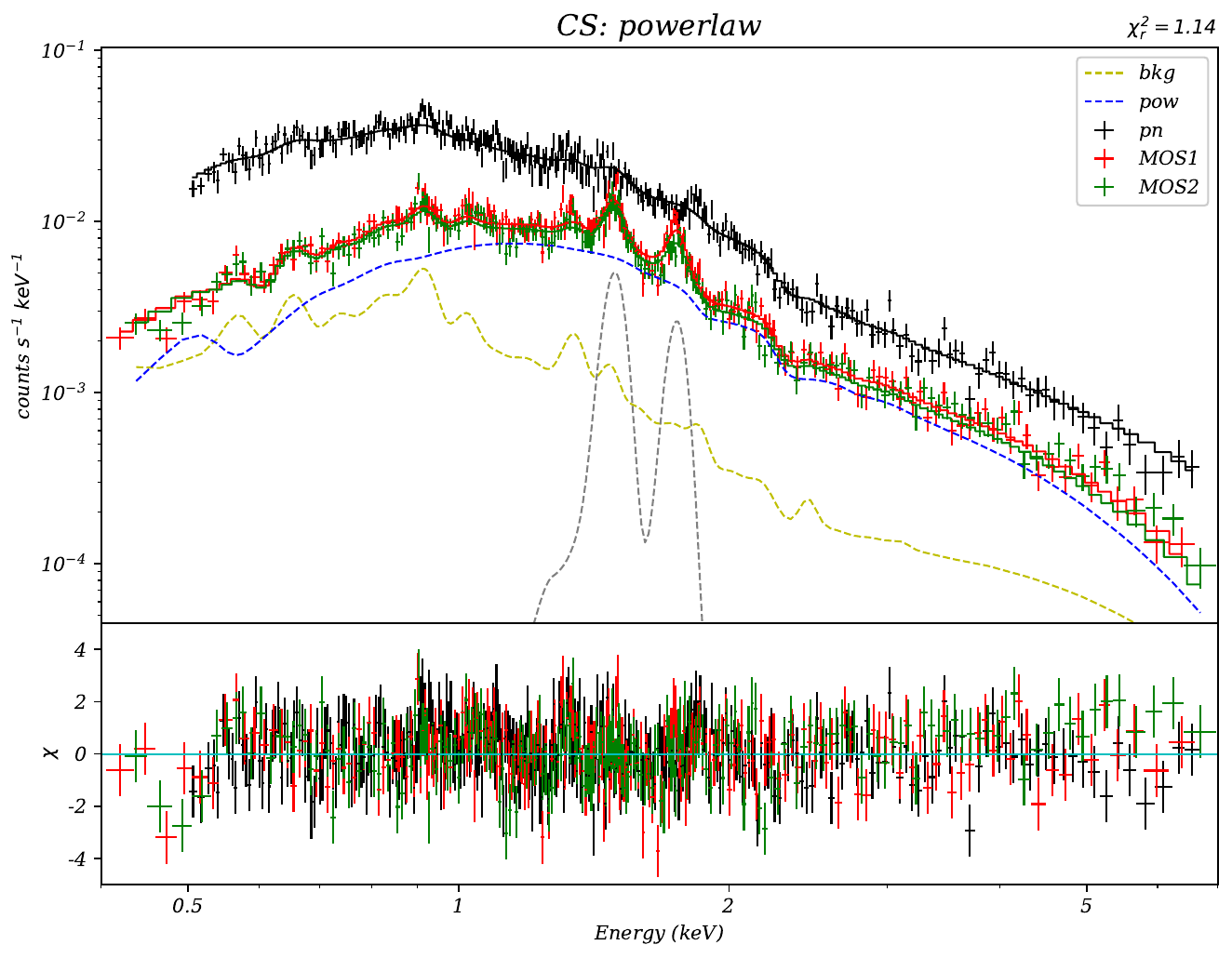}}
    \subfigure{\includegraphics[width=0.45\textwidth]{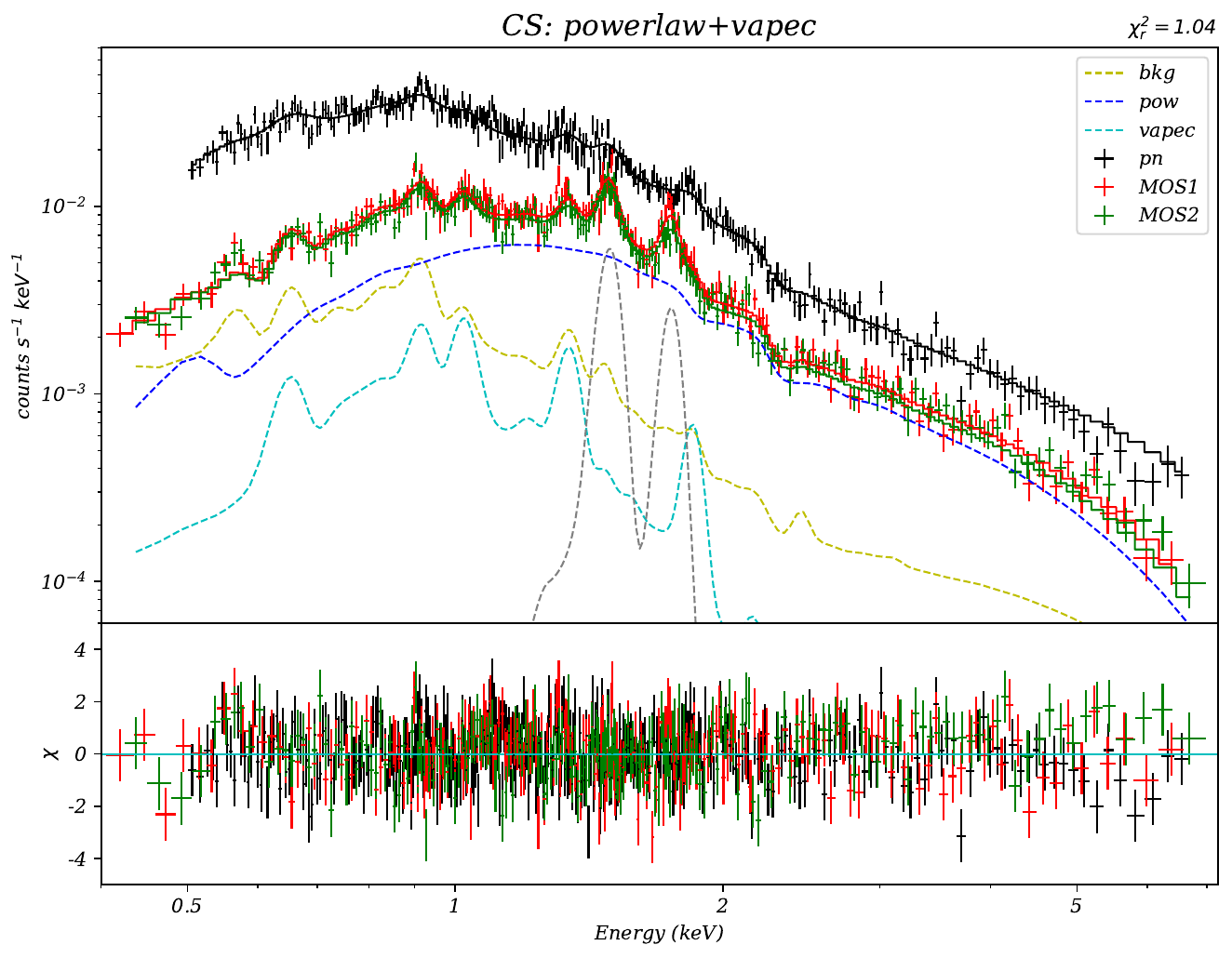}}
    \label{fig:CS}
    \caption{The spectra and the folded models of region CS.}
\end{figure*}

The spectra of different regions are presented here. Each region is fitted by a single \it powerlaw \rm model (left) or added with a \it vapec \rm model (right) as the thermal component. In each picture, we outline the different components of the model of MOS2 spectra with dash lines, including astrophysical background (yellow), instrumental Al and Si K-$\alpha$ lines (grey), non-thermal emission (\it powerlaw\rm, blue) and thermal emission (\it vapec\rm, cyan). The reduced chi-squared is labeled on the upper right of each image. The spectra of regions NW1, NW2, W, SW, CN, and CS are plotted in Fig. \ref{fig:NW1}1, \ref{fig:NW2}2, \ref{fig:W}3, \ref{fig:SW}4, \ref{fig:CN}5, and \ref{fig:CS}6, respectively.


\section{Winds of the O stars}
\label{sec:ostar}
Apart from 8 WR stars, \citet{Testor1993} listed 25 O-type stars in the OB association LH\,90 powering SB \sbd. For lack of direct measurement of their mass loss rates and wind velocity, we apply a relationship between the stellar parameters and stellar winds \citep{Kudritzki2000}. The mass loss rate $\Dot{M}$ can be expressed as:
\begin{equation}
    \Dot{M}=\frac{D_{\rm mom}}{v_w}\left(\frac{R_*}{R_{\sun}}\right)^{-\frac{1}{2}},
\end{equation}
where $D_{\rm mom}$ is the modified wind momentum, $v_w$ the wind velocity, and $R_*$ the star radius. Here, the subscript $_{\sun}$ represents the corresponding solar value, and the subscript $*$ refers to quantities of stars. Adopting
\begin{equation}
    \frac{L_*}{L_{\sun}}=\left(\frac{R_*}{R_{\sun}}\right)^2\left(\frac{T_*}{T_{\sun}}\right)^4
\end{equation}
and
\begin{equation}
    \lg D_{\rm mom}=\lg D_0+\xi\lg\frac{L_*}{L_{\sun}}
\end{equation}
given by \citet{Kudritzki2000}, the mass loss rate can be rewritten as the function of the stellar luminosity $L_*$ and effective temperature $T_*$:
\begin{equation}
    \Dot{M}=\frac{D_0}{v_w}\left(\frac{L_*}{L_{\sun}}\right)^{\xi-\frac{1}{4}}\left(\frac{T_*}{T_{\sun}}\right),
\end{equation}
where the accurate value of $D_0$, $\xi$, and $v_w$ are adpoted from \citet{Kudritzki2000} in terms of different luminous classes (LCs). Then we use bolometric magnitude $M_{\rm bol}$ to substitute for the luminosity $L_*$:
\begin{equation}
    M_{\sun}-M_{\rm bol}=\frac{5}{2}\frac{L_*}{L_{\sun}}.
\end{equation}
Finally, the mass loss rate is
\begin{equation}
    \Dot{M}=\left(4.00\times10^{-12}\times3.37^{-M_{\rm bol}}\,M_{\sun}\,{\rm yr}^{-1}\right)\left(\frac{T_*}{T_{\sun}}\right)\left(\frac{v_w}{1000\,\mathrm{km\,s^{-1}}}\right)^{-1}
\end{equation}
for main-sequence stars and giants (LC=V and III, respectively), and
\begin{equation}
   \Dot{M}=\left(2.03\times10^{-11}\times3.19^{-M_{\rm bol}}\,M_{\sun}\,{\rm yr}^{-1}\right)\left(\frac{T_*}{T_{\sun}}\right)\left(\frac{v_w}{1000\,\mathrm{km\,s^{-1}}}\right)^{-1}
\end{equation}
for supergiants (LC=I). The mechanical luminosity $L_w=\frac{1}{2}\Dot{M}v_w^2$ is
\begin{equation}
    L_w=\left(1.27\times10^{30}\times3.37^{-M_{\rm bol}}\,\mathrm{erg\,s^{-1}}\right)\left(\frac{T_*}{T_{\sun}}\right)\left(\frac{v_w}{1000\,\mathrm{km\,s^{-1}}}\right)
\end{equation}
for main-sequence stars and giants, and
\begin{equation}
    L_w=\left(1.28\times10^{31}\times3.19^{-M_{\rm bol}}\,\mathrm{erg\,s^{-1}}\right)\left(\frac{T_*}{T_{\sun}}\right)\left(\frac{v_w}{1000\,\mathrm{km\,s^{-1}}}\right)
\end{equation}
in terms of supergiants. \citet{Testor1993} lists the LCs, effective temperatures, and bolometric magnitudes of 25 O-type stars in LH\,90, which are shown in Table \ref{tab:owind} along with the mass loss rates and mechanical luminosities. Stars without clear LCs are considered as main-sequence stars or giants here.

\begin{table*}
    \centering
    \caption{The mechanical luminosities of 25 O stars in LH~90}
    \label{tab:owind}
    \begin{threeparttable}
    \begin{tabular}{ccccccc}
    \hline
        ID$^a$ & Sp-type$^a$ & $M_{\rm bol}^a$ & $\lg T_*^a$ & $v_w^b$ & $\lg\Dot{M}$ & $L_w$ \\
        ~ & ~ & (mag) & ~ & (km\,s$^{-1}$) & ~ & ($10^{36}$\,erg\,s$^{-1}$) \\ \hline
        2-01 & O5.5 III & -9.73  & 4.62  & 2000 & -5.71  & 2.49  \\ 
        2-02 & O7: & -8.07  & 4.60  & 2400 & -6.68  & 0.38  \\ 
        2-03 & O5.5 V & -9.61  & 4.64  & 2000 & -5.75  & 2.26  \\ 
        2-04 & O8 V & -7.43  & 4.58  & 1900 & -6.94  & 0.13  \\ 
        2-08 & O9.7 I & -10.63  & 4.48  & 1700 & -4.85  & 25.89  \\ 
        2-10 & O3-6 & -7.76  & 4.65  & 2900 & -6.88  & 0.35  \\ 
        2-11 & O9 V & -7.39  & 4.55  & 1500 & -6.89  & 0.09  \\ 
        2-13 & O6: f & -9.21  & 4.60  & 2300 & -6.06  & 1.46  \\ 
        2-14 & O5.5 III & -7.86  & 4.62  & 2000 & -6.69  & 0.26  \\ 
        2-16 & O8 V & -7.26  & 4.58  & 1900 & -7.03  & 0.11  \\ 
        2-17 & O8 V & -7.98  & 4.58  & 1900 & -6.65  & 0.26  \\ 
        2-20 & O5.5 III & -8.95  & 4.62  & 2000 & -6.12  & 0.97  \\ 
        2-22 & O3 III(f*) & -9.91  & 4.67  & 3000 & -5.74  & 5.22  \\ 
        2-24 & O6 V & -8.09  & 4.63  & 2600 & -6.68  & 0.45  \\ 
        2-26 & O9 V & -7.22  & 4.55  & 1500 & -6.98  & 0.08  \\ 
        2-28 & O8 V & -8.54  & 4.58  & 1900 & -6.35  & 0.51  \\ 
        2-29 & O9: & -8.21  & 4.53  & 1500 & -6.47  & 0.24  \\ 
        2-30 & O7: & -9.41  & 4.58  & 2400 & -6.00  & 1.85  \\ 
        2-32 & O6: I & -9.41  & 4.58  & 2300 & -5.50  & 10.71  \\ 
        2-33 & O5.5 III & -9.25  & 4.62  & 2000 & -5.96  & 1.39  \\ 
        2-36 & O8 V & -9.13  & 4.58  & 1900 & -6.04  & 1.04  \\ 
        2-38A & O9-B0 & -9.34  & 4.48  & 1500 & -5.93  & 0.84  \\ 
        2-50 & O7 III & -8.96  & 4.58  & 2600 & -6.27  & 1.16  \\ 
        2-51 & O7 V & -8.20  & 4.60  & 2400 & -6.61  & 0.45  \\ 
        2-53 & O6 If & -10.33  & 4.58  & 2300 & -5.03  & 31.13  \\ \hline
        Total & ~ & ~ & ~ & ~ & ~ & 89.71 \\ \hline
    \end{tabular}

    \begin{tablenotes}
        \item
        \textbf{Notes: }The units of $T$ and $\Dot{M}$ are K and $\mathrm{M_{\sun}}$\,yr$^{-1}$, respectively. This result requires a further correction for low metallicity in the LMC. \\
        \textbf{Reference: }$^a$\citet{Testor1993}; $^b$\citet{Kudritzki2000}
    \end{tablenotes}
    \end{threeparttable}
\end{table*}

Noticeably, three supergiants contribute over three-fourths of the total power from the O stars, of which 2-08 and 2-32 are in or near the sub-cluster associated with X-ray sources 12 and 6, respectively. The star 2-53, also named Sk~-69~212, has the highest mechanical luminosity and spatially corresponds to the soft X-ray source 4. The power of a supergiant here is comparable to that of a WR star.

However, this result requires further correction due to the LMC's low metalicity. \citet{Puls1996} suggested a difference of 0.20 dex in the wind momentum between the Galaxy and the LMC, and the wind velocity in the LMC is slightly slower \citep[Fig. 10 therein]{Kudritzki2000}. Hence, we roughly estimate the total mechanism luminosity of O stars at $\sim5\times10^{37}\rm\,erg\,s^{-1}$.


\bsp	
\label{lastpage}
\end{document}